\documentclass[11pt, oneside]{article}

\usepackage[margin=1in]{geometry}
\usepackage{tikz}
\usepackage{graphicx}
\usepackage{amssymb}
\usepackage{pgfplots}
\usepackage[font=small,labelfont=bf]{caption}
\usepackage{subcaption}
\usepackage{algorithm}
\usepackage{algpseudocode}
\usepackage{amsmath}
\usepackage{booktabs}
\usepackage{array}
\usepackage[numbers,sort&compress]{natbib}
\usepackage[hidelinks]{hyperref}
\usepackage{pgfplots}
\usepackage{setspace}
\usepackage{multirow}
\usepackage{mathtools}
\usepackage[colorinlistoftodos, textsize=small]{todonotes}

\numberwithin{equation}{section}
\usepgfplotslibrary{external}
\usepgfplotslibrary{groupplots}
\usetikzlibrary{matrix, calc}
\pgfplotsset{compat=1.18, lua backend=false}

\usepgfplotslibrary{fillbetween}
\usetikzlibrary{intersections}

\definecolor{myblue}{RGB}{31, 119, 180}
\definecolor{myred}{RGB}{214, 39, 40}
\definecolor{mygreen}{RGB}{44, 160, 44}
\definecolor{myorange}{RGB}{255, 127, 14}

\renewcommand\p@subfigure{\thefigure}

\newif\ifgeneratefigureone
\generatefigureonefalse
\newif\ifgeneratefiguretwo
\generatefiguretwofalse
\newif\ifgeneratefigurethree
\generatefigurethreefalse
\newif\ifgeneratefigurefour
\generatefigurefourfalse
\newif\ifgeneratefigurefive
\generatefigurefivefalse
\newif\ifgeneratefiguresix
\generatefiguresixfalse
\newif\ifgeneratefigureseven
\generatefiguresevenfalse
\newif\ifgeneratefigureeight
\generatefigureeightfalse
\newif\ifgeneratefigurenine
\generatefigureninefalse
\newif\ifgeneratefigureten
\generatefiguretenfalse
\newif\ifgeneratefigureeleven
\generatefigureelevenfalse
\newif\ifgeneratefiguretwelve
\generatefiguretwelvefalse
\newif\ifgeneratefigurethirteen
\generatefigurethirteenfalse
\newif\ifgeneratefigurefourteen
\generatefigurefourteenfalse
\newif\ifgeneratefigurefifteen
\generatefigurefifteenfalse
\newif\ifgeneratefiguresixteen
\generatefiguresixteenfalse
\newif\ifgeneratefigureseventeen
\generatefigureseventeenfalse
\newif\ifgeneratefigureeighteen
\generatefigureeighteenfalse
\newif\ifgeneratefigurenineteen
\generatefigurenineteenfalse
\newif\ifgeneratefiguretwenty
\generatefiguretwentyfalse
\newif\ifgeneratefigureenergygap
\generatefigureenergygapfalse
\newif\ifgeneratefiguregroundstateenergyuv
\generatefiguregroundstateenergyuvfalse
\newif\ifgeneratefiguregroundstateenergyir
\generatefiguregroundstateenergyirfalse
\newif\ifgeneratefiguregroundstateenergy
\generatefiguregroundstateenergyfalse
\newif\ifgeneratefigureqscaling
\generatefigureqscalingfalse
\newif\ifgeneratefigurebnlongsequence
\generatefigurebnlongsequencefalse
\newif\ifgeneratefigurefullkrylovcomplexity
\generatefigurefullkrylovcomplexityfalse
\newif\ifgeneratefigurefullkryloventropy
\generatefigurefullkryloventropyfalse
\newif\ifgeneratefiguresaturationkrylovcomplexity
\generatefiguresaturationkrylovcomplexityfalse
\newif\ifgeneratefiguresaturationkryloventropy
\generatefiguresaturationkryloventropyfalse
\newif\ifgeneratefiguremeanrtilde
\generatefiguremeanrtildefalse
\newif\ifgeneratefiguredkversusepsilon
\generatefiguredkversusepsilonfalse
\newif\ifgeneratefigurealphaVKcombined
\generatefigurealphaVKcombinedfalse
\newif\ifgeneratefigurerstatmass
\generatefigurerstatmassfalse
\newif\ifgeneratefigurerstatuv
\generatefigurerstatuvfalse
\newif\ifgeneratefigurerstatundeformed
\generatefigurerstatundeformedfalse
\newif\ifgeneratefigurerstatcSYK
\generatefigurerstatcSYKfalse
\newif\ifgeneratefigurekrylovexponentcombined
\generatefigurekrylovexponentcombinedfalse
\newif\ifgeneratefigurerstatoperator
\generatefigurerstatoperatorfalse

\begin{document}

\pagenumbering{gobble}

\begin{titlepage}
    \begin{flushright}
        QMUL-PH-25-33\
    \end{flushright}
    \null
    \vspace*{\fill}
    \begin{center}
        {\LARGE Krylov Complexity of Supersymmetric SYK Models\par}
        \vspace{1cm}
        {\large James Chryssanthacopoulos, David Vegh\par}
        \vspace{0.25cm}
        {\small \it Centre for Theoretical Physics, Department of Physics and Astronomy \\
        Queen Mary University of London, 327 Mile End Road, London E1 4NS, UK\par}
        \vspace{0.25cm}
        {\small \textbf{Email}: \texttt{j.chryssanthacopoulos@qmul.ac.uk}, \texttt{d.vegh@qmul.ac.uk}}
        \vspace{0.5cm}
        \begin{abstract}
            \noindent We study the effect of supersymmetry breaking on Krylov complexity in the $\mathcal{N}=2$ SYK model
            under irrelevant and mass deformations of the Hamiltonian. The irrelevant deformation breaks $\mathcal{N}=2$
            supersymmetry down to $\mathcal{N}=1$, while the mass deformation breaks supersymmetry completely. Using Krylov
            subspace methods, we analyze the Lanczos sequence, Krylov dimension, complexity, and entropy of the undeformed
            model and both deformations at finite system size. Both deformations enlarge the Krylov space and raise
            the saturation complexity as the BPS degeneracy is lifted. For the system sizes explored, the irrelevant deformation
            drives the saturation complexity to roughly half the Krylov dimension, while the mass deformation decreases it
            over an intermediate range of deformation strengths as the system drifts toward integrability. These distinct
            behaviors reveal how the mechanism of supersymmetry breaking leaves an imprint on quantum complexity.
        \end{abstract}
    \end{center}
    \vspace*{\fill}
\end{titlepage}

\newpage

\pagenumbering{gobble}
\tableofcontents
\newpage

\pagenumbering{arabic}
\setcounter{page}{1}

\section{Introduction}

Krylov complexity has recently emerged as a powerful tool for studying quantum chaos in many-body quantum systems and
gravity \cite{Nandy_2024, Rabinovici_2025, Baiguera_2025}. Roughly, Krylov complexity measures how quickly an operator grows under generic
Hamiltonian evolution.  In a chaotic system, Krylov complexity typically grows exponentially at early times in the large system limit, providing a
probe of quantum chaos. The notion of Krylov complexity grew out of a need to understand quantum chaos beyond the semiclassical
limit. Semiclassically, chaos is often diagnosed through the Lyapunov exponent, which quantifies the exponential growth
of out-of-time-order correlators. The Lyapunov exponent is bounded by  $2\pi T$, with $T$ the temperature of the system,
a bound saturated by maximally chaotic systems such as black holes \cite{MSS_2016, Tsuji_2018}. Krylov complexity extends this to more
general settings, quantifying operator growth under generic, non-integrable Hamiltonian dynamics. When the Lyapunov exponent can be
defined, it is bounded by the exponential growth rate of Krylov complexity \cite{Parker_2019}.

The fundamental idea behind Krylov complexity, and Krylov subspace methods generally, is to identify a minimal
subspace in which an operator evolves \cite{Liesen_2012}. This subspace is constructed by means of the Lanczos algorithm \cite{Cullum_2002}.
In this algorithm, an initial operator is evolved using the Liouvillian, the generator of time evolution. After each step,
the overlap with all previous operators is subtracted in a Gram-Schmidt-like procedure. The resulting orthonormal basis is called the
Krylov basis. In addition to the Krylov basis, the Lanczos algorithm generates a sequence of coefficients, called the
Lanczos coefficients, which correspond to the size of the Krylov basis elements. In a chaotic system, the Lanczos coefficients
initially grow in a system-dependent way before entering a regime of linear growth at intermediate times. The coefficients then
plateau and finally decay \cite{Parker_2019}. The Lanczos sequence captures the complete system dynamics, from which
Krylov complexity can be computed.

One interesting model which has been used to study Krylov complexity is the Sachdev-Ye-Kitaev (SYK) model, a quantum
mechanical model of interacting fermions with random couplings \cite{Sachdev_1993, kitaev2015talks, Maldacena_2016, Polchinski_2016,
Garcia_Garcia_2016, Rosenhaus_2019, Berkooz_2024}. Because its Lyapunov exponent saturates the bound, the SYK model is
maximally chaotic and has been proposed as a simple, one-dimensional model of quantum black holes. At low energies, SYK
exhibits emergent conformal symmetry. In the soft sector, the model can be described holographically by Jackiw-Teitelboim (JT)
gravity, a two-dimensional gravitational theory in anti-de Sitter space \cite{Jensen_2016, Mertens_2023, Turiaci_2024, Saad_2019}.
The Lanczos sequence of the model was shown to exhibit initial linear growth characteristic of chaotic systems
\cite{Parker_2019}. Krylov complexity also grows exponentially at early times. When the number of interactions is large,
the exponential growth rate of Krylov complexity matches the Lyapunov exponent, again suggesting maximal chaos.

One major way to generalize the SYK model is to introduce supersymmetry, which can be used to study supersymmetric
black holes \cite{Fu_2017, Berkooz_2020}. There are different versions of supersymmetric SYK depending on the number of
supercharges. Roughly, supersymmetric SYK models possess the same form as their non-supersymmetric counterparts, but the
couplings are no longer independent. These correlations change the structure of the large-$N$ equations and the
conformal dimension of the fermions, but generally the models are maximally chaotic in the sense that their Lyapunov
exponents saturate the bound \cite{Peng_2017}. Starting at $\mathcal{N}=2$, which possesses two complex supercharges,
the model develops a large degeneracy of ground states with exactly zero energy, called BPS states \cite{Heydeman_2024, Heydeman_2023}.
Between the discrete BPS states and non-BPS spectrum, there is a gap of order $1/S$, where $S$ is the entropy. Because
of its interesting properties, $\mathcal{N}=2$ SYK has been used to study the thermodynamics of extremal, supersymmetric
black holes in higher dimensions.

Although the Lyapunov exponent of supersymmetric SYK in the large-$N$ limit has been studied extensively, Krylov complexity of
supersymmetric SYK has received less attention. A study of Krylov complexity of $\mathcal{N}=1$ SYK has been conducted,
but it was embedded within a larger study of $T\bar{T}$-deformed SYK that also analyzed other measures of chaos \cite{He_2022}.
The chaotic properties of $\mathcal{N}=2$ have also been studied, but primarily within the context of random matrix theory
\cite{Kravtsov_2012, Livan_2018, Kar_2022, Kanazawa_2017, Li_2017, Guhr_1997}. The related concept of spread complexity has also been applied
to supersymmetric systems, but not to $\mathcal{N}=2$ SYK \cite{Das2025, Balasubramanian_2022}. This paper is devoted to a
systematic finite-size study of Krylov complexity of the $\mathcal{N}=2$ SYK model. Various deformations are explored that break the
supersymmetry, which enable the effects of supersymmetry on Krylov complexity to be studied.

The first deformation that is treated is a UV deformation introduced in \cite{Heydeman_2024}. Because this
deformation is irrelevant, it does not spoil the IR physics, which corresponds to the gravitational dynamics. The
deformation breaks $\mathcal{N}=2$ down to $\mathcal{N}=1$, eliminating the BPS states and lifting the ground-state
degeneracy for even parametrically small values of the deformation parameter. This deformation can be seen as
interpolating between two types of black holes \cite{Iliesiu_2021, Heydeman_2021}. The first corresponds to extremal,
supersymmetric black holes which carry a large entropy reflecting the ground-state degeneracy. The second type of black
hole corresponds to near-extremal, non-supersymmetric black holes which display no ground-state degeneracy and an average
energy spacing that is exponentially suppressed in the entropy.

While the UV deformation is interesting on the level of black hole physics, this paper studies the effect of the
deformation on Krylov complexity at finite $N$ and infinite temperature. The primary focus is the late-time saturation
of Krylov complexity and Krylov entropy. The saturation behavior of the undeformed model is first studied in sectors mixing
BPS and non-BPS states, revealing how the BPS structure governs the late-time dynamics. The effect of the deformation on
the saturation complexity is then analyzed, tracking how the breaking of supersymmetry modifies this behavior. The analysis
is performed numerically, which presents significant computational challenges.  These challenges include the exponential
growth of the Hilbert space with system size and the breakdown of orthogonality in the Krylov basis as the number of iterations
increases. Addressing these challenges necessitates the use of optimized, parallelized implementations of the Lanczos algorithm.

The second deformation that is explored is a relevant deformation that corresponds on the gravity side to deformations
away from two-dimensional anti-de Sitter space. There are many types of relevant deformations that can be considered, but
one of the simplest involves adding the Hamiltonian of two-body SYK with complex fermions, which is a free, integrable
theory. This has been studied in the non-supersymmetric case in the large-$N$ limit \cite{Anninos_2023, Chapman_2024}.
This mass deformation provides another pattern of supersymmetry breaking that breaks $\mathcal{N}=2$ directly down to $\mathcal{N}=0$.
The BPS degeneracy is lifted, and the energies are no longer bounded from below by zero. The incorporation of an integrable
Hamiltonian helps steer the model toward integrability, which has a direct bearing on Krylov complexity that distinguishes
between chaotic and integrable dynamics \cite{Garcia_Garcia_2018, Rabinovici_2022, Rabinovici_2022_2}. The same kind of analysis
as before is performed.

The remainder of this paper is organized as follows. In Section \ref{sec:review_susy_syk}, supersymmetric SYK models
are reviewed, focusing on $\mathcal{N}=2$ and its symmetries. Section \ref{sec:review_krylov} reviews Krylov complexity,
including the Lanczos algorithm. In Section \ref{sec:numerical_results}, numerical results are presented for the Krylov
dimension, Lanczos sequence, and Krylov complexity and entropy for the different deformations and symmetry sectors.
Section \ref{sec:conclusion} concludes with a discussion of findings and suggestions for future work.

\section{Supersymmetric SYK Models}
\label{sec:review_susy_syk}

This section provides the necessary background for understanding supersymmetric SYK. Section \ref{subsec:review_n0_syk}
reviews non-supersymmetric, or $\mathcal{N}=0$, SYK, including its symmetries and spectral properties. Then Section
\ref{subsec:review_n12_susy_syk} reviews SYK models with $\mathcal{N}=1$ and $\mathcal{N}=2$ supersymmetry. Section
\ref{subsec:deformations} introduces the deformations of the models which are relevant for the rest of the paper.

\subsection{$\mathcal{N}=0$ SYK}
\label{subsec:review_n0_syk}

Before introducing supersymmetric SYK, it is useful to review the basic properties of the regular SYK model, or $\mathcal{N}=0$
SYK. The SYK model is a $(0+1)$-dimensional quantum mechanical model of $N$ fermions with random interactions. Each
of the fermions can interact with any of the others, but only $q$ interact at a time, with $q$ even. There are two
versions of $\mathcal{N}=0$ SYK, one with Majorana fermions and another with Dirac fermions. In the Majorana version, the
Hamiltonian is given by
\begin{equation}
    H = i^\frac{q}{2}\sum_{1\leq i_1 < \cdots < i_q\leq N}J_{i_1\ldots i_q}\psi_{i_1}\cdots\psi_{i_q},
\end{equation}
where $\psi_i$ satisfy the algebra $\{\psi_i, \psi_j\} = \delta_{ij}$. The couplings $J_{i_1\ldots i_q}$ are drawn from
a Gaussian distribution with zero mean and variance
\begin{equation}
    \label{eq:coupling_variance}
    \langle J_{i_1\ldots i_q}^2 \rangle = \frac{J^2(q-1)!}{N^{q-1}}.
\end{equation}
The parameter $J$ is a dimension-one parameter controlling the interaction strength. The various numerical factors and
powers of $N$ are introduced to simplify the large-$N$ limit. The factor $i^\frac{q}{2}$ is included to ensure the
Hamiltonian is Hermitian.

At finite $N$, one way to study the SYK model is to construct the Hamiltonian explicitly in some basis and diagonalize
it. For even $N$, the model can be implemented using $N/2$ Dirac fermions $c_i$. These Dirac fermions map to Majorana
fermions in pairs:
\begin{equation}
    \label{eq:majorana-fermions}
    \psi_{2i-1} = \frac{i(c_i - c_i^\dagger)}{\sqrt{2}}, \quad
    \psi_{2i} = \frac{c_i + c_i^\dagger}{\sqrt{2}},
\end{equation}
where $\{c_i, c_j^\dagger\} = \delta_{ij}$ and $\{c_i, c_j\} = \{c_i^\dagger, c_j^\dagger\} = 0$. In practice, one way
to implement Dirac fermions is by means of the Jordan-Wigner transformation, which maps them to a chain of spin-1/2 particles. The construction is
\begin{equation}
    \label{eq:dirac-fermions}
    c_i = \prod_{k=1}^{i-1}(-\sigma_k^z)\frac{\sigma_i^x - i\sigma_i^y}{2}, \quad
    c_i^\dagger = \prod_{k=1}^{i-1}(-\sigma_k^z)\frac{\sigma_i^x + i\sigma_i^y}{2},
\end{equation}
where $\sigma_i^j$ is the Pauli matrix in the $j$ direction acting on the $i$th spin. The dimension of the Hilbert space
is $2^{N/2}$.

The total fermion number can be defined as $F \coloneq \sum_{i=1}^{N/2}c_i^\dagger c_i$. This is not a conserved charge, but
fermion number parity is conserved, $[H, (-1)^F] = 0$. This entails that the Hamiltonian can be block-diagonalized into sectors of
even and odd fermion number \cite{Cotler_2017}. There is also an antiunitary particle-hole operator, which maps $c_i \to c_i^\dagger$
and vice versa. It is given by
\begin{equation}
    P = K \prod_{i=1}^{N/2}(c_i + c_i^\dagger),
\end{equation}
where $K$ denotes complex conjugation. For $q = 0 \text{ mod } 4$, which includes the most common case $q=4$, $P$ commutes with the Hamiltonian.
For $N \text{ mod } 8 = 0$, $P^2 = 1$ and the two parity sectors separately exhibit Gaussian orthogonal ensemble (GOE) statistics. For $N \text{ mod } 8 = 4$,
$P^2 = -1$ and the parity sectors separately exhibit Gaussian symplectic ensemble (GSE) statistics. In this case, because
$P$ is an antiunitary operator that commutes with $H$, there is a two-fold Kramers degeneracy in each parity sector. For all other values of
$N$, the statistics follow the Gaussian unitary ensemble (GUE), and the energy levels are two-fold degenerate across the two parity sectors
\cite{Kanazawa_2017, Garcia_Garcia_2016, Garcia_Garcia_2017}.

The $\mathcal{N}=0$ SYK model involving Dirac fermions, called complex SYK, is built in much the same way as the Majorana version
\cite{Gu_2020, Pethybridge_2024, Zhang_2025, Sachdev_2015}. This model can be defined for any $q$-body interaction, but
for $q=4$ the Hamiltonian takes the form
\begin{equation}
    H = \sum_{ijkl} J_{ij;kl} \, c_i^\dagger c_j^\dagger c_k c_l,
\end{equation}
where $c_i$ can be constructed from Equation \eqref{eq:dirac-fermions}. The random couplings $J_{ij;kl}$ have the same
statistics as in the Majorana case, but are complex-valued. The indices are divided into two sets, obeying the relations
\begin{equation}
    J_{ji;kl} = J_{ij;lk} = -J_{ij;kl}, \quad \quad J_{kl;ij} = J_{ij;kl}^*.
\end{equation}
Because the Hamiltonian is built from complex instead of real fermions, the dimension of the Hilbert space is $2^N$
instead of $2^{N/2}$. In the remainder of this paper, unless otherwise stated, $\mathcal{N}=0$ SYK will refer to the complex SYK model.

\begin{figure}[bt]
    \centering
    \ifgeneratefigurerstatcSYK
        \tikzsetnextfilename{new_figure_r_stat_cSYK_combined}
        \input{figures/r_stat_cSYK_N12_combined.tex}
    \else
        \includegraphics{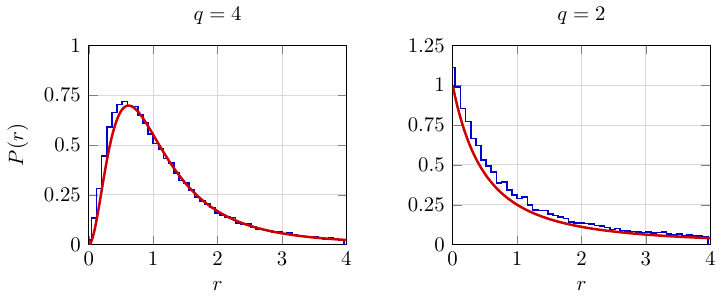}
    \fi
    \caption{The probability distribution of the ratio of level spacings $P(r)$ in the $\mathcal{N}=0$ SYK model. The left panel
    corresponds to the $q=4$ model, which exhibits GUE statistics, shown in red. The right panel corresponds to the $q=2$ model,
    which exhibits Poisson statistics. Each plot corresponds to 100 realizations of the Hamiltonian with $N=12$, $f=6$, and $J=1$.}
    \label{fig:r_stat_cSYK}
\end{figure}

The complex SYK model conserves fermion number, $[H, F] = 0$. Each fermion-number sector $f$ is non-degenerate, except $f=1$,
which has degeneracy $N$ from the $N$ single-fermion states. The model does not have a particle-hole symmetry and exhibits
GUE level statistics for all $N$. A standard diagnostic of chaos is the consecutive level-spacing ratio
\begin{equation}
    r \coloneqq \frac{\lambda_{i+2} - \lambda_{i+1}}{\lambda_{i+1} - \lambda_i},
\end{equation}
where $\lambda_i$ are the ordered energy eigenvalues. Since $F$ is a unitary symmetry, the statistics within each sector are
independent. Mixing sectors would artificially produce Poisson statistics, so the ratios must be computed sector by sector.
The left panel of Figure~\ref{fig:r_stat_cSYK} shows the probability distribution $P(r)$ for $\mathcal{N}=0$ SYK with $q=4$,
computed from 100 Hamiltonian realizations with $N=12$ and $f=6$. The Wigner-Dyson prediction is
\begin{equation}
    P_\text{WD}(r) = \frac{1}{Z_\beta}\frac{(r + r^2)^\beta}{(1 + r + r^2)^{1+3\beta/2}},
\end{equation}
where $\beta=1,2,4$ for GOE, GUE, and GSE, respectively, and $Z_1 = 8/27$, $Z_2 = 4\pi/81\sqrt{3}$, and $Z_4 = 4\pi/729\sqrt{3}$ \cite{Atas_2013}.
The numerical distribution agrees closely with the GUE prediction, shown in red, which is consistent with chaotic dynamics and the presence of eigenvalue repulsion.

A special case of the $\mathcal{N}=0$ SYK model is when $q=2$. In this case, the Hamiltonian contains only quadratic terms
of the form
\begin{equation}
    H = \frac{1}{2}\sum_{ij} J_{ij} \, c_i^\dagger c_j,
\end{equation}
where $J_{ij}$ is a Hermitian matrix whose elements are sampled from a Gaussian distribution with zero mean and variance
$J^2/N$. This model is a free, integrable theory, which can again be seen by studying the ratio of consecutive level spacings.
The right panel of Figure~\ref{fig:r_stat_cSYK} shows $P(r)$ for the $q=2$ model. The distribution follows the Poisson prediction
$P_\text{P}(r) = 1/(1+r)^2$, consistent with integrable dynamics and the absence of eigenvalue repulsion.

\subsection{$\mathcal{N}=1,2$ SYK}
\label{subsec:review_n12_susy_syk}

The SYK model can be generalized to include supersymmetry, which can be used to study supersymmetric black holes
\cite{Fu_2017, Peng_2017, Kanazawa_2017, Peng_2021}. There are various types of supersymmetric SYK models corresponding to
different numbers of supercharges, but two of the most studied are $\mathcal{N}=1$ and $\mathcal{N}=2$. The $\mathcal{N}=1$
model uses Majorana fermions, while $\mathcal{N}=2$ uses Dirac fermions. In $\mathcal{N}=1$, the supercharge $Q$
is defined as
\begin{equation}
    \label{eq:supercharge}
    Q = i^{\frac{q-1}{2}}\sum_{1\leq i_1 < \cdots < i_q\leq N}C_{i_1i_2\cdots i_q} \psi_{i_1} \psi_{i_2} \cdots \psi_{i_q},
\end{equation}
where $\psi_i$ are Majorana fermions and $q$ is taken to be odd. The variance of the couplings $C_{i_1\ldots i_q}$ is
the same as in Equation \eqref{eq:coupling_variance}. The supercharge is the generator of supersymmetric
transformations. The supersymmetric partners of the fermions $\psi_i$ are bosons $b_i$, which are not dynamical.
The Hamiltonian is the square of the supercharge, which simplifies to
\begin{equation}
    \label{eq:N_1_SYK_Hamiltonian}
    H = Q^2 = E_0 + \sum_{1 \leq i_1<\cdots <i_{\tilde{q}}\leq N} J_{i_1\cdots i_{\tilde{q}}} \psi_{i_1}\cdots \psi_{i_{\tilde{q}}},
\end{equation}
where $\tilde{q} = 2q-2$ and $E_0$ is the ground state energy. At finite $N$, the $\mathcal{N}=1$ model has a non-zero
ground state energy, which breaks supersymmetry, but it is recovered in the large-$N$ limit \cite{Fu_2017}. Notice that
the Hamiltonian of Equation \eqref{eq:N_1_SYK_Hamiltonian} has the same form as the Hamiltonian of $\mathcal{N}=0$ Majorana
SYK with $\tilde{q}$ interactions, but the couplings $J_{i_1\cdots i_{\tilde{q}}}$ are not independent. Instead, they roughly
correspond to the covariance matrix of the couplings $C_{i_1\cdots i_q}$.

One important symmetry of the $\mathcal{N}=1$ model absent in $\mathcal{N}=0$ is chiral symmetry in which the supercharge
anticommutes with the fermion parity operator, $\{Q, (-1)^F\} = 0$ \cite{Kanazawa_2017}. This property induces a block structure for $Q$:
\begin{equation}
    Q = \begin{pmatrix}
        0 & A \\
        A^\dagger & 0
    \end{pmatrix}.
\end{equation}
If $\lambda$ is an eigenvalue of $Q$, so is $-\lambda$. Because $H = Q^2$, the eigenvalues of $H$ are
at least two-fold degenerate. Each degenerate pair consists of one state from the even and one from the odd fermion parity sector.

The second major type of supersymmetric SYK model is the $\mathcal{N}=2$ model, which possesses two supercharges, $Q$ and $Q^\dagger$.
The Hamiltonian is given by
\begin{equation}
    H = \{Q, Q^\dagger\},
\end{equation}
which has the same form as the Hamiltonian of complex SYK with $\tilde{q}$ interactions, but with couplings that
are not fully independent. The supercharges are nilpotent, $Q^2 = {Q^\dagger}^2 = 0$, which follows from the
anticommutation relation between fermions, $\{c_i, c_j\} = 0$. Because of nilpotency, the Hamiltonian can also be written
$H = (Q + Q^\dagger)^2 = \mathcal{Q}^2$, where $\mathcal{Q} \coloneq Q + Q^\dagger$ is the self-adjoint supercharge.
The nilpotency of the supercharges also implies supersymmetry:
\begin{equation}
    \left[Q, H\right] = \big[Q, \{Q, Q^\dagger\}\big] = \big[Q^2, Q^\dagger\big] = 0\quad\text{if }\quad Q^2 = 0.
\end{equation}
Similarly, $[Q^\dagger, H] = 0$. To summarize, the fermionic anticommutation relations imply nilpotency of the
supercharges, which implements supersymmetry.

Like the complex SYK model, the $\mathcal{N}=2$ model conserves fermion number and fermion parity, but also admits
a particle-hole symmetry $P$. For $N \text{ mod } 4 = 0, 1$, $P^2 = 1$, while for $N \text{ mod } 4 = 2, 3$, $P^2 = -1$ \cite{Kanazawa_2017}.
Similar to the $\mathcal{N}=1$ model, the self-adjoint supercharge $\mathcal{Q}$ exhibits a chiral symmetry $\{\mathcal{Q}, (-1)^F\} = 0$,
so the even and odd fermion parity sectors are degenerate. The $\mathcal{N}=2$ model also possesses a $U(1)$ $R$-symmetry related to $F$,
generated by the operator
\begin{equation}
    R = \frac{1}{2q}\sum_i [c_i^\dagger, c_i] = \frac{F}{q} - \frac{N}{2q}.
\end{equation}
With this definition, the supercharges carry charge $\pm 1$, while the fermions carry charge $\pm 1/q$. The commutation
relations are
\begin{equation}
    [R, Q] = -Q,\quad [R, Q^\dagger] = Q^\dagger,\quad [R, c_i] = -\frac{1}{q}c_i,\quad [R, c_i^\dagger] = \frac{1}{q}c_i^\dagger.
\end{equation}
While the $U(1)_R$ charge does not commute with the supercharges, there is a $Z_q$ subgroup of this symmetry that does,
which transforms the fermions as $c_i \to e^{\frac{2\pi i n}{q}}c_i$, $n \in \mathbb{Z}_q$ \cite{Fu_2017}. The various
symmetries of the $\mathcal{N}=2$ model are summarized in Table \ref{tab:symmetries}.

\newcommand{\yes}{\textcolor{green!60!black}{Yes}}
\newcommand{\no}{\textcolor{red!70!black}{No}}

\begin{table}[bt]
\centering
\caption{Symmetry properties of the $\mathcal{N}=2$ SYK model and its deformations. The UV deformation breaks $\mathcal{N}=2$
down to $\mathcal{N}=1$, while the mass deformation breaks $\mathcal{N}=2$ down to $\mathcal{N}=0$. The UV deformation preserves
chiral symmetry, while the mass deformation preserves fermion number.}
\label{tab:symmetries}
\begin{tabular}{llccc}
\toprule
Symmetry & Constraint & Undeformed & UV deformed & Mass deformed \\
\midrule
Supersymmetry & $[H, Q] = 0 $ & $\mathcal{N}=2$ & $\mathcal{N}=1$ & $\mathcal{N}=0$ \\
Fermion parity & $[H,\,(-1)^F]=0$ & \yes & \yes & \yes \\
Fermion number & $[H,\,F]=0$ & \yes & \no & \yes \\
Chirality & $\{\mathcal{Q},\,(-1)^F\}=0$ & \yes & \yes & \no \\
Particle-hole & $[H,\,P]=0$ & \yes & \yes & \no \\
\bottomrule
\end{tabular}
\end{table}

Because the $U(1)_R$ charge commutes with the Hamiltonian, the Hilbert space can be decomposed as $\mathcal{H} =
\bigoplus_{k} \mathcal{H}_k$, where $\mathcal{H}_k$ is the subspace of states with $U(1)_R$ charge $k$.
In each space $\mathcal{H}_k$, there are BPS and non-BPS supermultiplets. BPS states are zero-energy states annihilated
by the supercharges $Q$ and $Q^\dagger$ \cite{Chen2024, Chang2024}. Non-BPS states are positive-energy states raised or lowered to other $R$-charge
sectors by the supercharges. The BPS states are highly degenerate, with a number that grows exponentially with $N$. For
$q=3$, the number of BPS states is \cite{Kanazawa_2017}
\begin{equation}
    \label{eq:n_bps}
    N_\text{BPS} = \begin{cases}
    4\cdot 3^{N/2-1}, & N \text{ mod } 4 = 0,2, \\
    2\cdot 3^{(N-1)/2}, & N \text{ mod } 4 = 1,3. \\
    \end{cases}
\end{equation}
For $N \text{ mod } 4 = 1$, there are additional exceptional BPS states not captured by this formula and which do not scale with
$N$. There are at least two such states for every $N \text{ mod } 4 = 1$, with exact diagonalization revealing six at $N=9$
\cite{Kanazawa_2017}.

A proper analysis of the symmetry structure of the $\mathcal{N}=2$ model requires going beyond the symmetries of $\mathcal{Q}$
and involves a decomposition of the eigenspaces of $F$. Using the fact that $H$, $QQ^\dagger$, $Q^\dagger Q$, and $F$
mutually commute, together with the nilpotency of the supercharges, each fermion number sector $\mathcal{H}_f$ decomposes as \cite{Kanazawa_2017}
\begin{equation}
    \mathcal{H}_f = \mathcal{H}_f^+ \oplus \mathcal{H}_f^- \oplus \mathcal{H}_f^z,
\end{equation}
where $\mathcal{H}_f^+$ is spanned by states annihilated by $Q$ but not $Q^\dagger$, $\mathcal{H}_f^-$ by states annihilated by
$Q^\dagger$ but not $Q$, and $\mathcal{H}_f^z$ is the BPS sector of states annihilated by both. The level statistics of the non-BPS sectors
$\mathcal{H}_f^\pm$ depend on $N$ and $f$. Consider the case of $q=3$ for concreteness, but similar patterns
hold for other values of $q$. For $N \text{ mod } 4 = 0, 2$, both $\mathcal{H}_f^+$ and $\mathcal{H}_f^-$
exhibit GUE statistics for all $f$. For $N \text{ mod } 4 = 1$, $\mathcal{H}_f^+$ is GSE at $f = (N - 3)/2$, and GUE otherwise,
while $\mathcal{H}_f^-$ is GSE at $f = (N + 3)/2$, and GUE otherwise. For $N \text{ mod } 4 = 3$, $\mathcal{H}_f^+$ and $\mathcal{H}_f^-$
are GOE at $f = (N - 3)/2$ and $f=(N + 3)/2$, respectively, and GUE otherwise \cite{Kanazawa_2017}. Crucially, this structure
relies on nilpotency of the supercharges and associated supersymmetry.

The degeneracy structure differs between the BPS and non-BPS sectors. All BPS states have exactly zero energy, so their
degeneracy equals the number of BPS states. In the non-BPS sector, the degeneracy depends on $N \text{ mod } 4$. For
$N \text{ mod } 4 = 0, 1, 2$, there is a four-fold degeneracy consisting of states with different fermion number \cite{Kanazawa_2017}
\begin{equation}
    \psi \in \mathcal{H}_f^+, \quad Q^\dagger \psi \in \mathcal{H}_{f+3}^-, \quad P\psi \in \mathcal{H}_{N-f}^-, \quad PQ^\dagger \psi \in \mathcal{H}_{N-f-3}^+,
\end{equation}
for $0 \leq f \leq N -3$. To describe the degeneracy structure for $N \text{ mod } 4 = 3$, let $N_f^+$ and $N_f^-$ denote the
dimensions of $\mathcal{H}_f^+$ and $\mathcal{H}_f^-$, respectively, given by \cite{Kanazawa_2017}
\begin{align}
    \label{eq:N_f_formulas}
    N_f^+ &= (-1)^{f+f_0}N_{f_0}^+ + (-1)^f \sum_{n=1}^{(f-f_0)/3} (-1)^{3n + f_0}\binom{N}{f_0 + 3n}, \\
    N_f^- &= \binom{N}{f} - N_f^+,
\end{align}
where $f_0 \coloneq f - 3\lfloor f/3\rfloor \in \{0,1,2\}$ and $N_0^+ = 1$, $N_1^+=N$, and $N_2^+ = N(N-1)/2$. The GOE
sectors at $f = (N - 3)/2$ and $f = (N + 3)/2$ contain $N_{(N-3)/2}^+ = N_{(N+3)/2}^-$ doublets, while the remaining GUE
sectors consist of quadruplets. The level statistics and degeneracy structure of the non-BPS sector are summarized in Table~\ref{tab:rmt}.

\begin{table}[bt]
\centering
\caption{Random matrix class and energy degeneracy $d_E$ in the non-BPS sector of the $\mathcal{N}=2$ SYK model and its deformations. The
non-BPS sector of the undeformed model falls into $f^\pm$ classes, which are GSE, GOE, or GUE depending on $N$ and $f$. The
UV deformation alters the level statistics while preserving a residual degeneracy. The mass deformation modifies the statistics to GUE
or Poisson, but lifts the degeneracy.}
\label{tab:rmt}
\begin{tabular}{ccccccc}
\toprule
& \multicolumn{2}{c}{Undeformed}
& \multicolumn{2}{c}{UV deformed}
& \multicolumn{2}{c}{Mass deformed} \\
\cmidrule(lr){2-3}\cmidrule(lr){4-5}\cmidrule(lr){6-7}
$N \bmod 4$ & Class & $d_E$ & Class & $d_E$ & RMT & $d_E$ \\
\midrule
$0$ & GUE      & $4$       & GOE & $2$ & GUE/Poisson & $1$ \\
$1$ & GSE/GUE  & $4$       & GSE   & $4$ & GUE/Poisson & $1$ \\
$2$ & GUE      & $4$       & GSE & $4$ & GUE/Poisson & $1$ \\
$3$ & GOE/GUE  & $2$ or $4$& GOE & $2$ & GUE/Poisson & $1$ \\
\bottomrule
\end{tabular}
\end{table}

\begin{figure}[bt]
    \centering
    \ifgeneratefigurerstatundeformed
        \tikzsetnextfilename{new_figure_r_stat_undeformed}
        \input{figures/r_stat_undeformed_combined.tex}
    \else
        \includegraphics{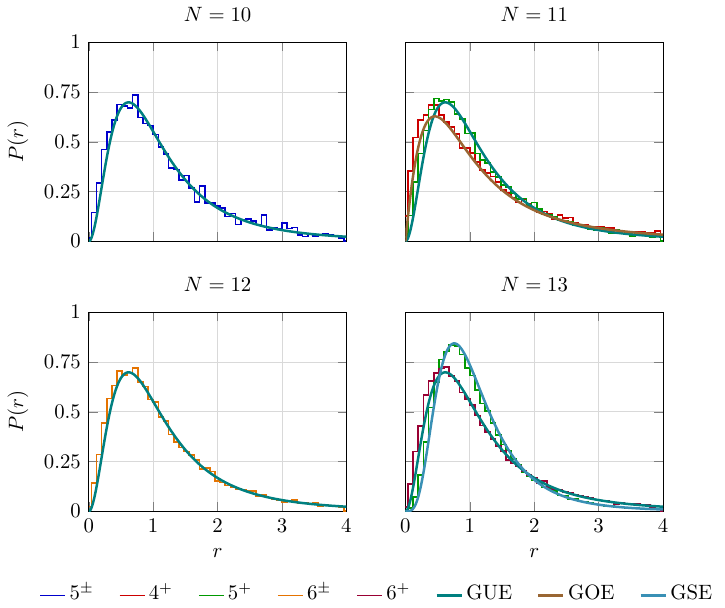}
    \fi
    \caption{The probability distribution of the ratio of level spacings $P(r)$ in the non-BPS sector of $\mathcal{N}=2$ SYK model.
    When $N \text{ mod } 4 = 0,2$, the statistics are GUE for all $f$. When $N \text{ mod } 4 = 1,3$, the statistics
    are either GSE or GOE when $f$ takes on a special value, and GUE otherwise. Each plot corresponds to 100 realizations
    of the Hamiltonian with $q=3$ and $J=1$.}
    \label{fig:r_stat_susy}
\end{figure}

To study spectral correlations, $\mathcal{H}_f^+$ and $\mathcal{H}_f^-$ must be treated separately, as they are statistically
independent. Figure~\ref{fig:r_stat_susy} shows $P(r)$ in the non-BPS sector for various values of $N$ and $f$, alongside the
GOE, GSE, and GUE predictions. In each case, the distribution is consistent with the predicted level statistics given the
values of $f$ and $N \text{ mod } 4$.

\subsection{Deformations}
\label{subsec:deformations}

It is possible to study deformations of the supersymmetric SYK model that break supersymmetry. Generally, supersymmetry
can be broken by adding any term to the Hamiltonian that is not a square of a supercharge. This paper explores two types
of deformations. The first is an irrelevant deformation that does not modify the conformal phase of the theory at low energies,
which corresponds to the gravitational sector. The second is a mass deformation which alters the gravitational dynamics
and more drastically modifies the chaotic properties of the model.

\emph{UV deformation.} The first deformation can be implemented by adding UV terms to the Lagrangian. Taken from \cite{Heydeman_2024}, the
deformation is given by
\begin{equation}
    \label{eq:irrelevant_deformed_Lagrangian}
    \mathcal{L} \rightarrow \mathcal{L} + \epsilon\left(i\sum_i c_i \partial_\tau c_i - \sum_i b_i^2 + \text{h.c.}\right),
\end{equation}
where $\textnormal{h.c.}$ denotes the Hermitian conjugate of the terms in parentheses, and $\epsilon$ controls the strength of the
deformation. One can show that this Lagrangian leads to a Hamiltonian with the same form as before, with the supercharge $Q$
having the same form as in Equation \eqref{eq:supercharge}, provided the fermions satisfy an $\epsilon$-deformed algebra:
\begin{equation}
    \label{eq:epsilon_deformed_fermion_algebra}
    \{c_i, c^\dagger_j\} = \frac{1}{1 - \epsilon^2}\delta_{ij},\quad
    \{c_i, c_j\} = -\{c^\dagger_i, c^\dagger_j\} = \frac{i\epsilon}{1-\epsilon^2}\delta_{ij}.
\end{equation}
When $\epsilon=\pm 1$, these expressions are singular, due to the presence of fermionic and bosonic zero modes.
This restricts $\epsilon$ to the range $|\epsilon| < 1$.

Starting from the fermions $c_i$ that satisfy the $\epsilon$-deformed algebra in Equation \eqref{eq:epsilon_deformed_fermion_algebra},
it is possible to introduce a change of basis to bring them into canonical form, given by \cite{Heydeman_2024}
\begin{equation}
    \label{eq:epsilon_deformed_fermion_transformation}
    c_i = \frac{\cos \frac{\tilde{\epsilon}}{2}}{\cos \tilde{\epsilon}}\chi_i +
    i\frac{\sin\frac{\tilde{\epsilon}}{2}}{\cos\tilde{\epsilon}}\chi^\dagger_i,
\end{equation}
where $\epsilon=\sin\tilde{\epsilon}$, with $|\tilde{\epsilon}| < \pi/2$. This naturally restricts to
$|\epsilon| < 1$. If $c_i$ and $c^\dagger_i$ satisfy the $\epsilon$-deformed algebra, then $\chi_i$ and
$\chi^\dagger_i$ satisfy the canonical fermionic algebra. When $\epsilon\rightarrow 0$, or equivalently
$\tilde{\epsilon}\rightarrow 0$, $c_i\rightarrow\chi_i$ and the model reduces to the original $\mathcal{N}=2$ model.
This transformation also provides a prescription for constructing the Hamiltonian starting from canonical fermions. One
can generate canonical fermions, apply the transformation of Equation \eqref{eq:epsilon_deformed_fermion_transformation},
then insert the resulting fermions into the supercharge. This demonstrates that it is not possible to pick a basis such
that $Q$ is holomorphic in the fermion fields and, simultaneously, the fermions satisfy the canonical algebra.

\begin{figure}[bt]
    \centering
    \ifgeneratefiguregroundstateenergy
        \tikzsetnextfilename{new_figure_ground_state_energy}
        \input{figures/e0_combined.tex}
    \else
        \includegraphics{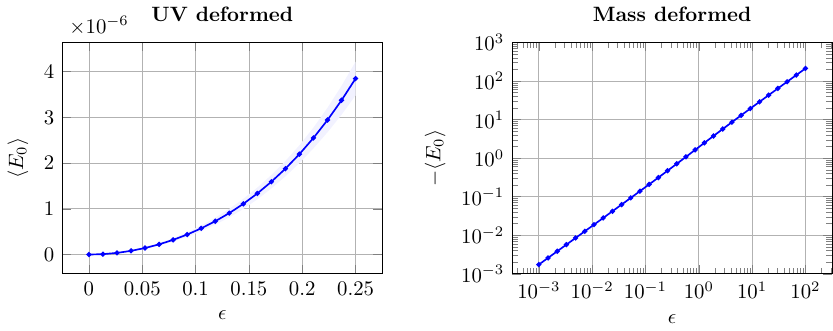}
    \fi
    \caption{The average ground state energy $E_0$ as a function of $\epsilon$ for $N=10$. In the UV-deformed model, the ground state energy
is lifted to a positive value, and is well approximated by $E_0 \sim \epsilon^2$. In the mass-deformed model with $f=5$, the
ground state energy becomes negative and is given by $E_0 \sim -2\epsilon$. Each plot was computed using $50$ realizations
of the Hamiltonian.}
    \label{fig:ground_state_energy}
\end{figure}

\begin{figure}[bt]
    \centering
    \ifgeneratefigureenergygap
        \tikzsetnextfilename{new_figure_energy_gap}
        \input{figures/gap_vs_epsilon_combined.tex}
    \else
        \includegraphics{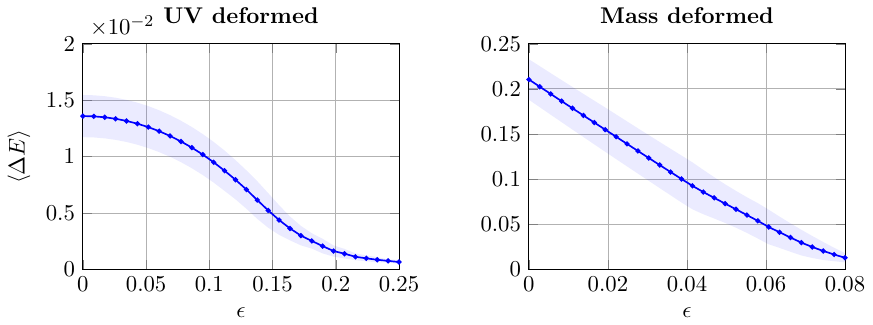}
    \fi
    \caption{The average energy gap between the ground state and the first excited state as $\epsilon$ varies for the
    UV and mass deformations for $N=10$. The mass-deformed model was restricted to the fermion number sector $f=5$.
    The gap was computed using $5{,}000$ realizations of the Hamiltonian, divided into $10$ groups to compute the standard deviation,
    shown as the shaded region.  The values $q=3$ and $J=1$ were used.}
    \label{fig:energy_gap}
\end{figure}

The UV deformation has a significant effect on the energy spectrum. In the undeformed model, the ground state energy is
exactly zero and a gap separates the BPS states from the non-BPS spectrum. The deformation lifts the ground state energy
to a positive value and closes this gap. The left panel of Figure~\ref{fig:ground_state_energy} shows the ground state energy
$E_0$ as a function of $\epsilon$ for $N=10$, which is well approximated by $E_0 \sim \epsilon^2$ \cite{Heydeman_2024}.
Figure~\ref{fig:energy_gap} shows the energy gap between the ground state and the first excited state with varying $\epsilon$.
Since the deformation preserves the number of would-be BPS states below the gap, the gap is computed as the energy difference
between the highest would-be BPS state and the lowest non-BPS state. As noted in \cite{Heydeman_2024}, because the non-BPS
spectrum is relatively sparse at small $N$, a large number of Hamiltonian samples is required to ensure the mean level
spacing is well below the gap scale. The gap narrows continuously with increasing $\epsilon$, closing around
$\epsilon \approx 0.25$.

The UV deformation breaks $\mathcal{N}=2$ supersymmetry down to $\mathcal{N}=1$. The deformed fermionic algebra $\{c_i, c_j\} \neq 0$
renders the supercharges no longer nilpotent, destroying the $\mathcal{H}_f^\pm$ decomposition. Since $F$ is no longer conserved,
$f$ is no longer a good quantum number and the Hilbert space cannot be decomposed into fermion-number sectors. The self-adjoint supercharge
$\mathcal{Q}$ remains a symmetry of the Hamiltonian, so $\mathcal{N}=1$ supersymmetry is preserved along with the chiral symmetry
$\{\mathcal{Q}, (-1)^F\} = 0$. This ensures fermion parity is conserved, $[H, (-1)^F] = 0$, and that the spectrum retains a
two-fold degeneracy between the even and odd fermion parity sectors. The particle-hole symmetry $P$ is also preserved under the
deformation. The symmetries of the UV-deformed model are summarized in Table~\ref{tab:symmetries}.

The breaking of $\mathcal{N}=2$ to $\mathcal{N}=1$ supersymmetry changes the level statistics to those of the $\mathcal{N}=1$
model. For $q=3$, the model exhibits GOE statistics when $N \text{ mod } 4 = 0, 3$, and GSE statistics when
$N \text{ mod } 4 = 1, 2$ \cite{Kanazawa_2017, Li_2017}. The two-fold degeneracy from chiral symmetry is present throughout, with an additional Kramers degeneracy
when $N \text{ mod } 4 = 1, 2$, giving a four-fold degeneracy. Table~\ref{tab:rmt} summarizes the level statistics
and degeneracy structure of the UV-deformed model. For $N \text{ mod } 4 = 0$, the statistics switch from GUE to GOE and the degeneracy
is reduced from four to two. For $N \text{ mod } 4 = 2$, the statistics switch from GUE to GSE, and the Kramers degeneracy
ensures the four-fold degeneracy is maintained. For $N \text{ mod } 4 = 1$, the GSE statistics that previously occurred only
in special $f$ sectors become the statistics of the full model, with the degeneracy unchanged. For $N \text{ mod } 4 = 3$,
the GOE statistics of special $f$ sectors become the statistics of the full model, and the
two-fold degeneracy of the GOE doublets is preserved.

\begin{figure}[bt]
    \centering
    \ifgeneratefigurerstatuv
        \tikzsetnextfilename{new_figure_r_stat_uv}
        \input{figures/r_stat_uv_combined.tex}
    \else
        \includegraphics{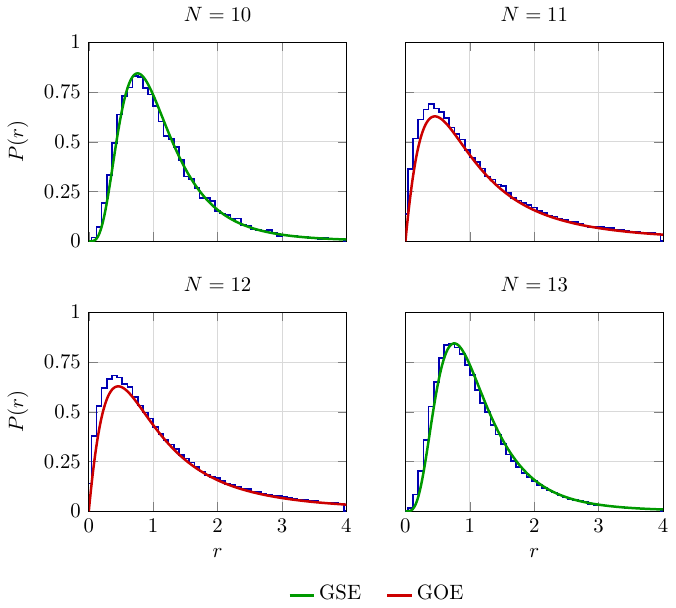}
    \fi
    \caption{The probability distribution of the ratio of level spacings $P(r)$ in the UV-deformed $\mathcal{N}=2$ SYK model.
    When $N \text{ mod } 4 = 0,3$, the statistics are GOE, while when $N \text{ mod } 4 = 1,2$, the statistics are GSE.
    Each plot corresponds to 100 realizations of the Hamiltonian in the even parity sector with $N=10$, $\epsilon=0.6$, and $J=1$.}
    \label{fig:r_stat_susy_uv}
\end{figure}

Figure~\ref{fig:r_stat_susy_uv} shows the distribution of $r$ for the UV-deformed model for various values of $N$, computed in the even fermion
parity sector at $\epsilon=0.6$, when the deformation has fully set in. For $N \text{ mod } 4 = 0,3$, the distribution is
consistent with GOE statistics, while for $N \text{ mod } 4 = 1,2$, it is consistent with GSE statistics.

\begin{figure}[bt]
    \centering
    \ifgeneratefigurerstatmass
        \tikzsetnextfilename{new_figure_r_stat_mass}
        \input{figures/r_stat_mass_combined.tex}
    \else
        \includegraphics{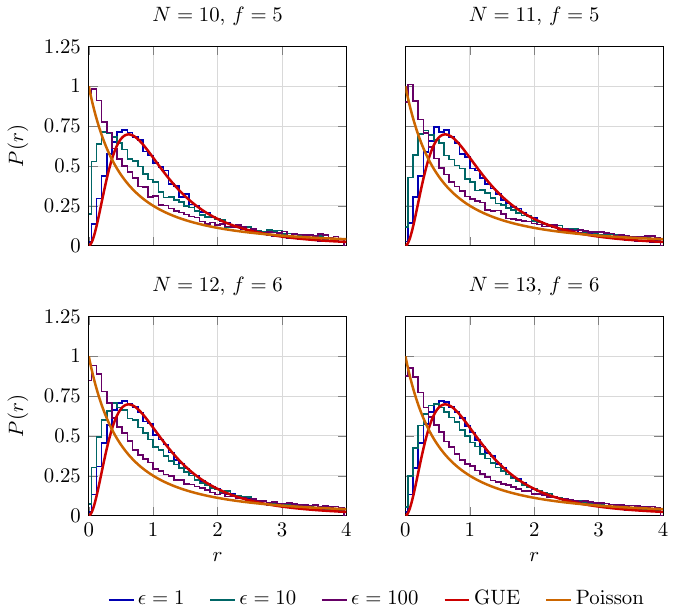}
    \fi
    \caption{The probability distribution of the ratio of level spacings $P(r)$ in the mass-deformed $\mathcal{N}=2$ SYK model.
    When the size of the deformation $\epsilon$ is small, the distributions are consistent with GUE statistics. As
    $\epsilon$ increases, the distribution smoothly approaches the Poisson distribution, which is indicative of integrable
    dynamics. Each plot corresponds to 100 realizations of the Hamiltonian with $J=1$.}
    \label{fig:r_stat_susy_ir}
\end{figure}

\emph{Mass deformation.} The second deformation that will be explored is a relevant deformation. Relevant deformations correspond to deformations
of the two-dimensional anti-de Sitter vacuum in the gravity picture. Generally, the non-supersymmetric SYK model admits
a set of relevant deformations given by
\begin{equation}
    H = H_{\mathcal{N}=0}^q + \sum_i \epsilon_i H_{\mathcal{N}=0}^{\tilde{q}_i}.
\end{equation}
The Hamiltonian is a sum of multiple Hamiltonians with different numbers of fermions, $q$ and $\tilde{q}_i$. If $\tilde{q}_i < q$,
the second term acts as a relevant deformation controlled by the dimensionless couplings $\epsilon_i$ \cite{Anninos_2023,
Chapman_2024, Hamdan2025}. This picture is modified slightly when moving to the $\mathcal{N}=2$ model. In this case, a class of
relevant deformations can be written
\begin{equation}
    H = H_{\mathcal{N}=2}^q + \sum_i \epsilon_i H_{\mathcal{N}=0}^{\tilde{q}_i}.
\end{equation}
This Hamiltonian is the sum of the $\mathcal{N}=2$ Hamiltonian with $q$ interactions and a set of $\mathcal{N}=0$ Hamiltonians
with $\tilde{q}_i$ interactions. In $\mathcal{N}=2$ SYK, the conformal dimension of the fermions is $1/2q$ instead of $1/q$.
This means the deformation is relevant if $\tilde{q}_i < 2q$. For the $q=3$ model considered in this paper, this restricts
$\tilde{q}_i$ to two or four. The deformation that will be considered involves a single term with $\tilde{q}_i=2$ given by
\begin{equation}
    \label{eq:relevant_deformed_Hamiltonian}
    H = H_{\mathcal{N}=2}^{q=3} + \frac{1}{2}\epsilon\sum_{ij} J_{ij} c_i^\dagger c_j.
\end{equation}
Because the second term in the Hamiltonian corresponds to a free theory, the deformation interpolates between a chaotic and integrable system.
Note that the $\epsilon$ used here is distinct from the one used in the irrelevant deformation. While $\epsilon$ of the irrelevant
deformation is constrained to lie inside $|\epsilon| < 1$, the $\epsilon$ of the mass deformation is valid for all $\epsilon \geq 0$.

The mass deformation breaks $\mathcal{N}=2$ supersymmetry completely, along with other symmetries such as chiral and particle-hole symmetries.
Since both terms in the mass-deformed Hamiltonian separately conserve fermion number, $F$ and fermion parity remain good quantum numbers.
The symmetries are summarized in Table \ref{tab:symmetries}. The level statistics interpolate between GUE at small $\epsilon$ and Poisson
at large $\epsilon$, reflecting a transition from chaotic to integrable dynamics. The degeneracy structure of the $\mathcal{N}=2$ model
is fully broken, leaving a non-degenerate spectrum. These properties are summarized in Table~\ref{tab:rmt}. Since the Hamiltonian is no
longer positive-definite, the ground state energy is negative. The right panel of Figure~\ref{fig:ground_state_energy} shows the ground state energy
as a function of $\epsilon$ for $N=10$ and $f=5$, which is well approximated by $E_0 \sim -2\epsilon$. The right panel of
Figure~\ref{fig:energy_gap} shows the energy gap as a function of $\epsilon$ for the mass deformation. The gap decreases roughly linearly,
closing around $\epsilon \approx 0.08$.

Figure~\ref{fig:r_stat_susy_ir} shows the distribution of $r$ for the mass-deformed model across various fermion-number sectors
and values of $\epsilon$. At small $\epsilon$, the distribution is consistent with GUE statistics for all values of $N$ and $f$.
As $\epsilon$ increases, it transitions smoothly toward the Poisson distribution, signaling a crossover from chaotic to integrable dynamics.

\begin{figure}[bt]
    \centering
    \ifgeneratefiguremeanrtilde
        \tikzsetnextfilename{new_figure_mean_rtilde}
        \input{figures/mean_rtilde_vs_eps_combined.tex}
    \else
        \includegraphics{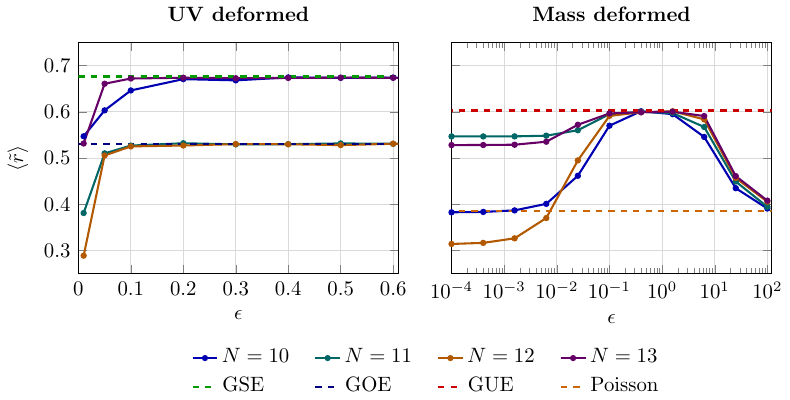}
    \fi
    \caption{Mean of the ratio of level spacings $\tilde{r}$ as the deformation strength $\epsilon$ varies.
    Under the UV deformation, the mean of $\tilde{r}$ transitions to the GOE or GSE value depending on $N \text{ mod } 4$, while
    under the mass deformation, the mean of $\tilde{r}$ transitions smoothly from the GUE value to the Poisson value.
    The UV model was diagonalized in the even parity sector, while the mass-deformed model was diagonalized in the $f=\lfloor N/2 \rfloor$ sector.
    Each plot was computed using $50$ realizations of the Hamiltonian.}
    \label{fig:mean_r_vs_eps}
\end{figure}

Figure~\ref{fig:mean_r_vs_eps} shows the mean of $\tilde{r} = \min(r, 1/r)$ as a function of $\epsilon$ for both deformations.
The left panel shows the UV-deformed model in the even fermion parity sector for various values of $N$. As $\epsilon$ increases,
$\langle \tilde{r} \rangle$ transitions to the GOE or GSE value depending on $N \text{ mod } 4$. The transition occurs by
$\epsilon \approx 0.2$, which coincides with the closing of the energy gap. The right panel shows the mass-deformed model in
the $f=\lfloor N/2 \rfloor$ sector for various values of $N$. At small $\epsilon$, $\langle \tilde{r} \rangle$ rises toward
the GUE value of approximately $0.60$, consistent with chaotic dynamics. As $\epsilon$ increases further, $\langle \tilde{r} \rangle$
decreases smoothly toward the Poisson value of $0.39$, signaling the crossover to integrable dynamics.

\section{Krylov Complexity and the SYK Model}
\label{sec:review_krylov}

This section reviews the main ideas behind Krylov complexity and its application to the SYK model. Section \ref{subsec:krylov_space}
introduces the notion of Krylov space and discusses how to construct it using the Lanczos algorithm. Then Section \ref{subsec:krylov_dimension}
discusses the size of the Krylov space, called the Krylov dimension, and describes how to compute it for $\mathcal{N}=2$ SYK
and its deformations. Section \ref{subsec:krylov_complexity_entropy} concludes with a review of Krylov complexity and entropy.

\subsection{Krylov Space}
\label{subsec:krylov_space}

Under time evolution by a Hamiltonian, a generic operator will trace out a trajectory in Hilbert space. In many
instances, this operator will not explore the entire Hilbert space, but instead will be confined to a subspace. Krylov
space is one way of defining this subspace in which the dynamics unfolds \cite{Nandy_2024}. To develop these ideas,
consider a Hermitian operator $\mathcal{O}$ evolving in time under the action of a time-independent Hamiltonian $H$.
In the Heisenberg picture, the dynamics of the operator is governed by the equation
\begin{equation}
    \partial_t \mathcal{O}(t) = i[H, \mathcal{O}(t)] = i\mathcal{L}\mathcal{O}(t).
\end{equation}
The operator $\mathcal{L}$ is called the Liouvillian superoperator. Its action on an operator is given by the commutator
of that operator with the Hamiltonian, $\mathcal{L}\,\cdot \coloneq [H, \cdot]$. The solution to the Heisenberg equation is
\begin{equation}
    \mathcal{O}(t) = e^{i\mathcal{L}t}\mathcal{O}.
\end{equation}
In words, the operator at time $t$ is obtained by applying powers of the Liouvillian on the initial operator. This
evolution is contained in a subspace of operator space, called Krylov space, that is spanned by $\{\mathcal{L}^n\mathcal{O}\}_{n=0}^\infty =
\{\mathcal{O}, \mathcal{L}\mathcal{O}, \mathcal{L}^2\mathcal{O}, \ldots\}$.
The dimension of this Krylov space is the number of linearly independent operators generated in this way. As time evolves, the operator generally requires contributions from increasingly many Krylov basis elements.
In the infinite temperature setting discussed in this work, the Hilbert-Schmidt inner product can be used to
define products between operators in Krylov space, given by \cite{Nandy_2024}
\begin{equation}
    \label{eq:inner_product}
    (\mathcal{O}_1|\mathcal{O}_2) \coloneq \frac{1}{d}\mathrm{Tr}(\mathcal{O}_1^\dagger \mathcal{O}_2).
\end{equation}

Krylov space can be constructed using an algorithm called the Lanczos algorithm. This algorithm constructs an orthonormal basis
with respect to the inner product defined above using a Gram-Schmidt-like procedure. Starting from an initial operator $\mathcal{O}$,
the algorithm is as follows:
\begin{enumerate}
    \item Set $|\mathcal{O}_{-1})= 0$, $|\mathcal{O}_0) = |\mathcal{O})/\sqrt{(\mathcal{O}|\mathcal{O})}$, $b_0 = 0$, and $n=1$.
    \item Act with the Liouvillian to produce the next state, making sure to subtract off the previous state:
    $|\mathcal{A}_n) = \mathcal{L}|\mathcal{O}_{n-1}) - b_{n-1}|\mathcal{O}_{n-2})$.
    \item If $||\mathcal{A}_n|| = 0$, stop. Otherwise, set $b_n = ||\mathcal{A}_n||$ and $|\mathcal{O}_n) = |\mathcal{A}_n)/b_n$.
    \item Repeat the process, setting $n \to n+1$ and returning to step 2.
\end{enumerate}
The process terminates at $n = D_K$, the Krylov dimension. The output is a $D_K$-dimensional orthonormal ordered basis
$\left\{|O_n) \,\right\}_{n=0}^{D_K - 1}= \left\{|O_0),\, |O_1),\, \ldots,\, |O_{D_K - 1})\right\}$, called the
Krylov basis. While this basis is not Hermitian, it can be made Hermitian by multiplying each element $|O_n)$ by $i^n$.

In addition to the Krylov basis, another output of the Lanczos algorithm is a set of non-negative coefficients $\{b_n\}_{n=0}^{D_K - 1}$,
called the Lanczos coefficients, which completely determine the dynamics of the operator in the basis. In many applications,
the Lanczos algorithm is numerically unstable due to the accumulation of rounding errors. The effect is that after a few
iterations, orthogonality of the Krylov basis may be lost. To ensure orthogonality, techniques like full orthogonalization or
partial reorthogonalization can be used \cite{Rabinovici_2021}.

After the Krylov basis has been constructed, it can be used to expand the operator at time $t$:
\begin{equation}
    |\mathcal{O}(t)) = \sum_{n=0}^{D_K - 1} i^n \varphi_n(t) |\mathcal{O}_n).
\end{equation}
The real-valued functions $\varphi_n(t)$ are known as the Krylov-basis wavefunctions, which represent the probability
amplitude for the operator at $t$ to be in state $|\mathcal{O}_n)$. From the Heisenberg equation, it can be shown that
the wavefunctions satisfy the differential equation
\begin{equation}
    \label{eq:hopping_equation}
    \dot{\varphi}_n(t) = b_n \varphi_{n-1}(t) - b_{n+1} \varphi_{n+1}(t),
\end{equation}
subject to the initial condition $\varphi_{-1}(t) = 0$ and $\varphi_n(0) = \delta_{0n}$. The dot represents a time
derivative. This equation describes a single-particle hopping model, where a particle at site $n$ hops to site $n-1$ with
rate $b_n$ and to site $n+1$ with rate $b_{n+1}$. The evolution of the operator is reduced to a hopping problem
on a one-dimensional lattice, known as the Krylov lattice. The dynamics on the Krylov lattice are fully determined by the
Lanczos coefficients.

Generically, the behavior of the Lanczos coefficients can be divided into three distinct regimes \cite{Barbon_2019, Rabinovici_2021}.
In the first regime, called the Lanczos ascent, the Lanczos coefficients grow initially until $n \sim O(S)$, where $S$
is the thermodynamic entropy. In the second regime, called the Lanczos plateau, the coefficients plateau at $b_n \sim
O(\Lambda S)$, where $\Lambda$ is the energy scale.  In the final regime, called the Lanczos descent, the $b_n$ decrease
at exponentially large $n \sim O(e^S)$.

According to the operator growth hypothesis, in the Lanczos ascent regime, the Lanczos coefficients should grow
linearly in $n$ for a generic chaotic system in $d$ dimensions, with an additional logarithmic correction in
one-dimensional systems \cite{Parker_2019}.  Explicitly, the hypothesis states that for a chaotic system
\begin{equation}
    \label{eq:operator_growth_hypothesis}
    b_n =
    \begin{cases}
    \displaystyle \frac{A n}{\log n} + o\left(\frac{n}{\log n}\right), & d = 1, \\
    \alpha n + \beta + o(1), & d > 1,
    \end{cases}
\end{equation}
where $A$, $\alpha$, and $\beta$ are constants. While the coefficients in these expressions may change for different
initial operators, the pattern of linear growth should remain the same.

In general, the Lanczos coefficients need to be computed numerically, but they can be computed exactly for the SYK
model when the number of interactions $q$ is large. This can be done using the moments method, starting from the
autocorrelation function $C(t) \coloneq (\mathcal{O}(t)|\mathcal{O})$ \cite{Viswanath_2013}. To compute $C(t)$ in the
large-$q$ limit, an expansion in $1/q$ can be performed:
\begin{equation}
    C(t) = 1 + \frac{1}{q}g(t) + O(1/q^2).
\end{equation}
For $\mathcal{N}=0$ SYK at infinite temperature, $g(t)$ satisfies \cite{Nandy_2024, Parker_2019, Heydeman_2024, Roberts_2018}
\begin{equation}
    g''(t) = -2\mathcal{J}^2 e^{g(t)},\quad \mathcal{J}^2 \coloneq 2^{1-q}qJ^2.
\end{equation}
With the initial conditions $g(0) = g'(0) = 0$, the solution is
\begin{equation}
    C(t) = 1 + \frac{2}{q}\ln\operatorname{sech}(\mathcal{J}t) + O(1/q^2).
\end{equation}
From the autocorrelation function, the moments $\mu_{2n} = C^{(2n)}(0)$ can be computed. The Lanczos coefficients can then be
calculated from the moments using a recursive algorithm \cite{Viswanath_2013}. The result is \cite{Nandy_2024, Parker_2019}
\begin{equation}
    \label{eq:large_q_lanczos_coefficients}
    b_n =
    \begin{cases}
    \displaystyle \mathcal{J}\sqrt{2/q} + O(1/q), & n=1, \\
    \mathcal{J}\sqrt{n(n-1)} + O(1/q), & n > 1.
    \end{cases}
\end{equation}
Comparing to Equation \eqref{eq:operator_growth_hypothesis}, $\mathcal{N}=0$ SYK in the large-$q$ limit satisfies the
operator growth hypothesis with linear growth rate $\alpha = \mathcal{J}$.

Similar calculations can be performed for the $\mathcal{N}=2$ SYK model. In this case, the first correction $g(t)$
satisfies \cite{Heydeman_2024}
\begin{equation}
    g''(t) = -\mathcal{J}^2 e^{2g(t)},
\end{equation}
which results in the Lanczos sequence
\begin{equation}
    \label{eq:large_q_lanczos_coefficients_N_2}
    b_n =
    \begin{cases}
    \displaystyle \mathcal{J}\sqrt{1/q} + O(1/q), & n=1, \\
    \mathcal{J}\sqrt{n(n-1)} + O(1/q), & n > 1.
    \end{cases}
\end{equation}
The leading behavior in $n$ is unmodified, $\alpha = \mathcal{J}$. After the irrelevant deformation of Equation
\eqref{eq:irrelevant_deformed_Lagrangian}, the function $g(t)$ satisfies \cite{Heydeman_2024}
\begin{equation}
    g''(t) = -\gamma^2\mathcal{J}^2e^{2g(t)},\quad\gamma \coloneq \frac{1+\epsilon^2}{(1-\epsilon^2)^2}.
\end{equation}
The overall effect is simply to replace $\mathcal{J}$ with $\gamma\mathcal{J}$, resulting in the leading behavior $\alpha
= \gamma\mathcal{J}$, which varies with the deformation strength. If the mass deformation of Equation \eqref{eq:relevant_deformed_Hamiltonian}
is applied instead, the equation for $g(t)$ becomes
\begin{equation}
    g''(t) = -\mathcal{J}^2\left(e^{2g(t)} + \epsilon^2\right).
\end{equation}
This equation is equivalent to that of $\mathcal{N}=0$ SYK with $q$-body interactions deformed by a quadratic term, up
to an overall rescaling of the coupling $\mathcal{J} \to \mathcal{J}/2$. Although an analytical solution is not
known, the equation can be solved numerically or perturbatively in $\epsilon$ \cite{Chapman_2024, Garcia_Garcia_2018}.

\subsection{Krylov Dimension}
\label{subsec:krylov_dimension}

In Krylov subspace methods, an important role is played by the Krylov dimension $D_K$. Formally, $D_K$ is the cardinality of
the maximal set of linearly independent vectors in the set $\{\mathcal{L}^n\mathcal{O}\}_{n=0}^\infty$. This set can be
written as a Vandermonde matrix \cite{Rabinovici_2021}. First, express the operator $\mathcal{O}$ in the basis $|\omega_{ab}) \coloneq |E_a\rangle \langle E_b|$ formed
by the eigenstates of the Hamiltonian. Each application of the Liouvillian to $\mathcal{O}$ can be expressed as a vector
in this basis, and the resulting vectors can be assembled into a matrix. The result is

\begin{equation}
    \label{eq:vandermonde_matrix}
    \begin{pmatrix}
        O_{11} & O_{22} & \cdots & O_{dd} & O_{12} & O_{13} & \cdots & O_{d{-}1,d} \\
        0 & 0 & \cdots & 0 & O_{12} \omega_{12} & O_{13} \omega_{13} & \cdots & O_{d{-}1,d} \omega_{d{-}1,d} \\
        0 & 0 & \cdots & 0 & O_{12} \omega_{12}^2 & O_{13} \omega_{13}^2 & \cdots & O_{d{-}1,d} \omega_{d{-}1,d}^2 \\
        \vdots & \vdots & \ddots & \vdots & \vdots & \vdots & \ddots & \vdots \\
        0 & 0 & \cdots & 0 & O_{12} \omega_{12}^{d^2 - 1} & O_{13} \omega_{13}^{d^2 - 1} & \cdots & O_{d{-}1,d} \omega_{d{-}1,d}^{d^2 - 1}
    \end{pmatrix}.
\end{equation}
Since the first $d$ columns of this matrix are linearly dependent, $d-1$ of them must be removed. This gives an upper
bound on the Krylov dimension \cite{Rabinovici_2021}:
\begin{equation}
    \label{eq:krylov_dimension_bound}
    D_K^\text{max} = d(d - 1) + 1.
\end{equation}

The bound above applies to any generic Hamiltonian, but can be tightened when the Hamiltonian has degeneracies. In the
presence of degeneracies, certain phases vanish, and the corresponding columns of the matrix must be removed. Letting
$d_E$ denote the average energy degeneracy, there are only $d/d_E$ distinct energy levels, and the bound becomes
\begin{equation}
    \label{eq:krylov_dimension_bound_degeneracy}
    D_K^\text{max} = \frac{d}{d_E}\left(\frac{d}{d_E} - 1\right) + 1.
\end{equation}
When $d_E = 1$, this reduces to Equation~\eqref{eq:krylov_dimension_bound}. Note that $D_K^\text{max}$ is always an integer,
since $d_E = d/n_E$, where $n_E$ is the number of distinct energy levels, so the bound can equivalently be written
$D_K^\text{max} = n_E(n_E - 1) + 1$. These bounds assume no accidental coincidences among the non-zero frequencies.

In $\mathcal{N}=2$ SYK, the BPS sector has a large ground-state degeneracy. The entire BPS subspace contributes a single
energy level, and the total number of distinct energy levels is $n_E = n^\pm + 1$, where $n^\pm$ is the number of distinct
non-BPS energy levels across $\mathcal{H}^+$ and $\mathcal{H}^-$. Then the Krylov dimension in the full spectrum is bounded by
\begin{equation}
    D_K^\text{max} = n^\pm(n^\pm + 1) + 1.
\end{equation}
One can estimate $n^\pm$ more explicitly. For $N \text{ mod } 4 = 0,1,2$, Table~\ref{tab:rmt} shows that the non-BPS
sector is exactly four-fold degenerate, giving $n^\pm = (d - N_\text{BPS})/4$, where $N_\text{BPS}$ is given by Equation~\eqref{eq:n_bps}.
For $N \text{ mod } 4 = 3$, the non-BPS sector contains $N_{(N-3)/2}^+$ doublets with the remaining $d - N_\text{BPS} - 2N_{(N-3)/2}^+$ states
forming quadruplets. The number of distinct non-BPS energy levels is then $n^\pm = (d - N_\text{BPS} + 2N_{(N-3)/2}^+)/4$, where $N_f^+$
is given by Equation \eqref{eq:N_f_formulas}.

After the UV deformation, the BPS degeneracy is lifted and only non-BPS states remain. The number of distinct non-BPS energy levels is
$d/d_E$, where $d_E$ can be read off from Table~\ref{tab:rmt}, giving
\begin{equation}
    D_K^\text{max} = \begin{cases}
    2^{N-1}\left(2^{N-1} - 1\right) + 1, & N \text{ mod } 4 = 0,3, \\
    2^{N-2}\left(2^{N-2} - 1\right) + 1, & N \text{ mod } 4 = 1,2. \\
    \end{cases}
\end{equation}
The mass-deformed model is non-degenerate, so $D_K^\text{max} = 2^N(2^N - 1) + 1$ for all $N$, which is larger than the
UV bound in general.

The bounds above are theoretical predictions for $D_K^\text{max}$ derived from the degeneracy structure of the Hamiltonian.
It is also possible to estimate $D_K^\text{max}$ numerically by counting the number of distinct non-zero phases, plus one
to account for the matrix element between identical energy levels. These approaches furnish an upper bound that makes no assumption
about the choice of operator $\mathcal{O}$. Once $\mathcal{O}$ is specified, the actual Krylov dimension $D_K$ can be computed,
and is generically smaller than $D_K^\text{max}$. If $\mathcal{O}$ has vanishing projection onto some energy eigenspaces,
the corresponding columns of the Vandermonde matrix must be removed. More precisely, $D_K$ equals the number of eigenspaces of $\mathcal{L}$
onto which $\mathcal{O}$ has non-zero projection \cite{Rabinovici_2021}. The more degeneracies the Liouvillian has, the lower
the Krylov dimension. In Section~\ref{subsec:krylov_dimension_results}, $D_K^\text{max}$ is computed numerically to confirm the theoretical predictions,
and $D_K$ is also computed for a specific choice of $\mathcal{O}$ using the procedure described above.

\subsection{Krylov Complexity and Entropy}
\label{subsec:krylov_complexity_entropy}

An important measure of operator growth is known as Krylov complexity, which is defined as the average position of
an operator in the Krylov chain
\begin{equation}
    \label{eq:krylov_complexity}
    K(t) = \sum_{n=0}^{D_K - 1} n\left| \varphi_n(t) \right|^2.
\end{equation}
By definition, $K(t) \geq 0$ and it vanishes at the initial time, $K(0) = 0$. Krylov complexity grows as the operator
moves away from the origin of the Krylov lattice, reflecting the fact that the Krylov basis elements are more nonlocal as
the lattice index is increased.

Similar to the Lanczos coefficients, the behavior of Krylov complexity can generically be divided into different regimes \cite{Barbon_2019}.
In the first regime, which corresponds to the Lanczos ascent, Krylov complexity grows exponentially until the scrambling time,
$t_* \sim O(\log S)$, reaching a value of $K \sim O(S)$. In the next regime, the Lanczos plateau, Krylov complexity grows
linearly until the Heisenberg time, $t_H \sim O(e^S)$, reaching a value of $K \sim O(e^{S})$. In this regime, the Krylov
wavefunctions are uniformly distributed over the Krylov space, $|\varphi_n(t > t_H)|^2 \sim 1/D_K$. The corresponding
Krylov complexity is given by $K(t > t_H) \sim D_K/2$. Krylov complexity stays at this plateau value during the Lanczos descent.
In general, the plateau value of Krylov complexity can be computed from the Krylov basis elements $|\mathcal{O}_n)$ and the eigenvectors
of the Liouvillian $|\omega_i)$ \cite{Rabinovici_2022}:
\begin{equation}
    K_\text{sat} = \sum_{n=0}^{D_K - 1} n\,\overline{|\varphi_n|^2}
    = \sum_{n=0}^{D_K - 1} \sum_{i=0}^{D_K-1} n\,{|(\mathcal{O}_n|\omega_i)|}^2\,{|(\omega_i|\mathcal{O}_0)|}^2 .
    \label{eq:ksat_diagonal_ensemble}
\end{equation}

In addition to Krylov complexity, Krylov entropy also captures how randomized the distribution of Krylov wavefunctions
is \cite{Barbon_2019}. It is given by
\begin{equation}
    \label{eq:krylov_entropy}
    S(t) = - \sum_{n=0}^{D_K - 1} \left| \varphi_n(t) \right|^2 \log \left| \varphi_n(t) \right|^2.
\end{equation}
Like Krylov complexity, Krylov entropy increases during the ascent phase. In general, beyond the scrambling time, Krylov
complexity and entropy are related logarithmically: $S(t) \sim \log K(t)$ \cite{Barbon_2019}.

Table \ref{tab:krylov-behaviors} summarizes the growth behaviors of the Lanczos coefficients, Krylov complexity, and Krylov
entropy for different types of systems. The table distinguishes between chaotic, integrable, and bounded systems. In
chaotic systems, the Lanczos coefficients grow linearly, leading to exponential growth of Krylov complexity and linear growth of
Krylov entropy. In integrable systems, the Lanczos coefficients grow sublinearly, $\alpha n^\delta$, with $\delta < 1$.
This leads to sub-exponential growth of Krylov complexity and logarithmic growth of Krylov entropy. In bounded systems,
which also captures the dynamics of chaotic systems at late times, the Lanczos coefficients are constant, leading to linear
growth of Krylov complexity and logarithmic growth of Krylov entropy. This corresponds to $\delta = 0$ in the integrable case.

\begin{table}[t]
    \centering
    \small
    \caption{Comparison of growth behaviors for the Lanczos coefficients $b_n$, Krylov complexity $K(t)$, and Krylov entropy $S(t)$ for
    different system types. Chaotic systems feature linear growth of $b_n$, leading to exponential growth of $K(t)$ and linear growth
    of $S(t)$. Integrable systems exhibit sublinear growth of $b_n$, which leads to power-law behavior of $K(t)$. Bounded systems
    correspond to integrable systems with $\delta = 0$.}
    \label{tab:krylov-behaviors}
    \begin{tabular}{cccc}
    \toprule
    & \multicolumn{3}{c}{\emph{System Type}} \\
    \cmidrule(lr){2-4}
    Quantity & Chaotic & Integrable & Bounded \\
    \midrule
    $b_n$       & $\alpha n$                  & $\alpha n^{\delta}$                & $b$ \\
    $K(t)$      & $e^{2\alpha t}$             & $(\alpha t)^{\frac{1}{1 - \delta}}$ & $bt$ \\
    $S(t)$      & $2\alpha t$                 & $\log(\alpha t)$                    & $\log(2bt)$ \\
    \bottomrule
    \end{tabular}
\end{table}

For the SYK model, Krylov complexity can be computed analytically in the large-$q$ limit. In $\mathcal{N}=0$, the
Lanczos coefficients of Equation \eqref{eq:large_q_lanczos_coefficients} can be substituted into Equation \eqref{eq:hopping_equation}
to produce the Krylov-basis wavefunctions. Substituting the result into Equation \eqref{eq:krylov_complexity}, Krylov
complexity is given by \cite{Nandy_2024}
\begin{equation}
    K(t) = \frac{2}{q}\sinh^2(\mathcal{J}t) + O(1/q^2),
\end{equation}
which is quadratic at the earliest times, $K(t)\sim \tfrac{2}{q}(\mathcal{J}t)^2$, then crosses over to exponential growth
$K(t)\sim e^{2\mathcal{J}t}$ once $\mathcal{J}t\gtrsim 1$. The exponential rate $2\alpha = 2\mathcal{J}$ is the operator-growth
signature of chaos. For the $\mathcal{N}=2$ model, the Lanczos coefficients of Equation \eqref{eq:large_q_lanczos_coefficients_N_2} lead to
\begin{equation}
    K(t) = \frac{1}{q}\sinh^2(\mathcal{J}t) + O(1/q^2),
\end{equation}
which has the same intermediate-time exponential rate $2\mathcal{J}$ as $\mathcal{N}=0$. Under the irrelevant deformation, the growth rate is modified
to $2\alpha = 2\gamma\mathcal{J}$, where $\gamma \coloneq (1+\epsilon^2)/(1-\epsilon^2)^2$. At finite $N$, deviations from this
exponential behavior are expected, and the growth profile is better characterized by the power-law ansatz of Table~\ref{tab:krylov-behaviors}.

\section{Numerical Results}
\label{sec:numerical_results}

This section presents numerical results for Krylov complexity in the $\mathcal{N}=2$ SYK model and its deformations.
As discussed in Section~\ref{sec:review_krylov}, the Krylov space depends on the choice of initial operator $\mathcal{O}$,
and in practice one may wish to restrict to a particular symmetry sector of the Hamiltonian to isolate the behavior of Krylov
complexity more clearly. The question is whether the Krylov basis elements remain within the same symmetry sector. Suppose $[H, \mathcal{S}] = 0$
for some symmetry operator $\mathcal{S}$. Since the Lanczos algorithm builds successive states by applying the Liouvillian to the previous state,
it follows that $[H, \mathcal{O}]$ commutes with $\mathcal{S}$ whenever $\mathcal{O}$ does. Restricting to a particular symmetry
sector is then achieved by choosing the initial operator to lie within the charge-zero sector of that symmetry \cite{Rabinovici_2022}.

The initial operator is chosen to be the hopping operator between sites $N$ and $N-1$:
\begin{equation}
    \mathcal{O} = h_{N,N-1} \coloneq c_N^\dagger c_{N-1} + c_{N-1}^\dagger c_N,
\end{equation}
which is Hermitian and traceless. The tracelessness is important, as a non-zero trace would introduce a one-point function
contribution to the autocorrelation function, suppressing the saturation value of Krylov complexity without reflecting
any universal behavior. Working with a traceless operator ensures that the two-point function coincides with its connected part,
capturing the universal features of operator growth \cite{Rabinovici_2022}. For notational brevity, the hopping operator
will be referred to simply as $h$ in the discussion below.

Because the hopping operator raises and then lowers the fermion number by one, it commutes with fermion number, $[F, \mathcal{O}] = 0$,
and consequently also with fermion parity, $[(-1)^F, \mathcal{O}] = 0$, and $R$-symmetry, $[R, \mathcal{O}] = 0$. These symmetries
enable the study of Krylov complexity in the UV-deformed model in fixed parity sectors, and in the mass-deformed model in fixed fermion
number sectors. Computing Krylov complexity of the undeformed model is more subtle, since $h$ does not commute with the
supercharges and so mixes the $\mathcal{H}^+$, $\mathcal{H}^-$, and $\mathcal{H}^z$ sectors. This case is handled with more care in the discussion below.

In Section \ref{subsec:krylov_dimension_results}, results for the Krylov dimension are presented. Then Section
\ref{subsec:lanczos_coefficients_results} analyzes the Lanczos sequence for various deformations and sectors. Section
\ref{subsec:krylov_complexity_entropy_results} concludes with results for Krylov complexity and entropy. The code used
to generate the results in this section is available in the ancillary files at \url{https://arxiv.org/abs/2511.20769}.

\subsection{Krylov Dimension}
\label{subsec:krylov_dimension_results}

As discussed in Section~\ref{subsec:krylov_dimension}, the Krylov dimension $D_K$ for a specific operator $\mathcal{O}$ is
computed by counting the number of distinct non-zero phases $\omega_{ab}$ with $a\neq b$ for which $O_{ab} \neq 0$, plus one
if any diagonal element $O_{aa} \neq 0$. This can be done by diagonalizing the Hamiltonian and projecting $\mathcal{O}$ onto
the energy eigenbasis. The theoretical bounds derived in Section~\ref{subsec:krylov_dimension} are agnostic to the choice of
$\mathcal{O}$ and should hold in particular for the hopping operator, though the actual $D_K$ may be lower.
For the UV and mass deformations, the Hamiltonian depends on the deformation strength $\epsilon$, so the Krylov dimension will
also vary with $\epsilon$.

\begin{figure}[bt]
    \centering
    \ifgeneratefiguredkversusepsilon
        \tikzsetnextfilename{new_figure_dk_vs_epsilon}
        \input{figures/dk_vs_epsilon_combined.tex}
    \else
        \includegraphics{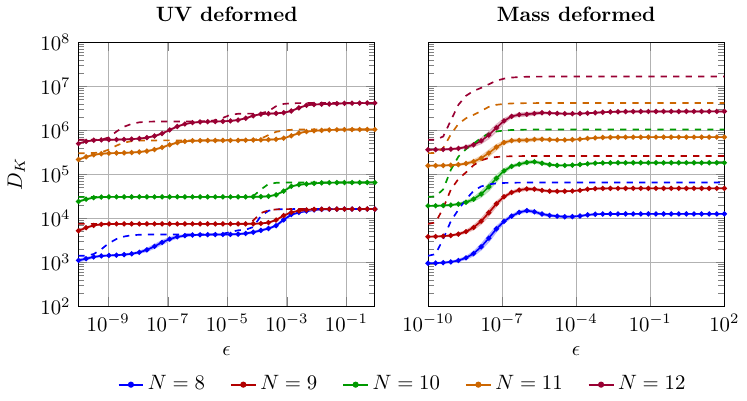}
    \fi
    \caption{The Krylov dimension $D_K$ for the hopping operator (solid) compared to the Krylov dimension bound (dashed)
    as the deformation strength $\epsilon$ varies for the UV and mass deformations. The tolerance for $\mathcal{O}_{ab}$ of the hopping operator in
    the energy basis to be considered zero was set to $10^{-10}$, while the tolerance for two energy levels to be
    considered degenerate was set to $10^{-8}$. Each point corresponds to the average of five samples, and the shaded regions
    show the standard deviation across samples.}
    \label{fig:dk_hopping_vs_epsilon}
\end{figure}

Figure~\ref{fig:dk_hopping_vs_epsilon} shows the Krylov dimension for the hopping operator alongside the bound as
the deformation strength $\epsilon$ is varied, for both the UV and mass deformations. The solid lines show $D_K$ for the hopping operator,
obtained via the procedure described above, and the dashed lines the empirical bound computed from the number of distinct phases
obtained by exact diagonalization. At very small $\epsilon$, the bound matches the theoretical predictions for the undeformed model in
Section~\ref{subsec:krylov_dimension}. As $\epsilon$ increases, the bound rises before plateauing at a value consistent with the theoretical
predictions for the deformed models. At large $\epsilon$, once the deformation has fully set in, the UV-deformed model saturates the
bound while the mass-deformed model remains well below it.

Table~\ref{tab:krylov_dimension_results} summarizes the Krylov dimension for the hopping operator alongside the theoretical bound
once the deformation has set in. These values are obtained from the energy-basis phase count described above. 
 In the $\mathcal{N}=2$ model, the large number of BPS states leads to high degeneracy, which suppresses
the Krylov bound. The Krylov dimension for the hopping operator falls moderately below the bound due to sparsity in the non-BPS sector,
though the operator has dense support in the BPS sector. On breaking to $\mathcal{N}=1$ by the UV deformation, the BPS degeneracy is
lifted and the Krylov bound increases. The residual degeneracy in the non-BPS sector, arising from chiral and other symmetries,
prevents the bound from reaching its maximum possible value. In this case, the hopping operator has non-zero projection onto all
distinct energy levels, so the Krylov dimension saturates the bound.

\begin{table}[bt]
\centering
\small
\setlength{\tabcolsep}{4pt}
\caption{The Krylov dimension for the hopping operator compared to the Krylov dimension
bound for the $\mathcal{N}=2$ SYK model and its deformations once they have fully set in. The Krylov dimension is moderately lower than the bound for the
undeformed model, saturates the bound for the UV-deformed model, and is significantly lower than the bound for the mass-deformed model.
Only the Krylov dimension for the undeformed model is subject to noise, while the numerical results for the other values match
the theoretical predictions well.}
\label{tab:krylov_dimension_results}
\begin{tabular}{rrrrrrrrr}
\toprule
& & \multicolumn{2}{c}{Undeformed}
& \multicolumn{2}{c}{UV deformed}
& \multicolumn{2}{c}{Mass deformed} \\
\cmidrule(lr){3-4}\cmidrule(lr){5-6}\cmidrule(lr){7-8}
$N$ & $d$ & $D_K^\text{max}$ & $D_K$ & $D_K^\text{max}$ & $D_K$ & $D_K^\text{max}$ & $D_K$ \\
\midrule
$8$  & $256$   & $1{,}407$        & $952$            & $16{,}257$       & $16{,}257$       & $65{,}281$        & $12{,}615$        \\
$9$  & $512$   & $7{,}483$        & $3{,}828$        & $16{,}257$       & $16{,}257$       & $261{,}633$       & $48{,}109$       \\
$10$ & $1{,}024$ & $30{,}801$     & $19{,}199$       & $65{,}281$       & $65{,}281$       & $1{,}047{,}553$   & $183{,}733$   \\
$11$ & $2{,}048$ & $303{,}051$     & $157{,}627$       & $1{,}047{,}553$       & $1{,}047{,}553$       & $4{,}192{,}257$   & $703{,}385$   \\
$12$ & $4{,}096$ & $610{,}743$     & $363{,}668$       & $4{,}192{,}257$       & $4{,}192{,}257$       & $16{,}773{,}121$   & $2{,}700{,}061$   \\
\bottomrule
\end{tabular}
\end{table}

Breaking to $\mathcal{N}=0$ by the mass deformation lifts the degeneracy entirely, and the Krylov bound increases significantly,
exceeding even that of the UV model. In the mass-deformed model the spectrum is non-degenerate, so every energy eigenstate has
definite fermion number $f$. Since $[F,h]=0$, the hopping operator is block-diagonal in $f$ in the energy eigenbasis, with
$h_{ab}=0$ unless $f_a=f_b$. The distinct non-zero Liouvillian frequencies available to $h$ therefore number at most
$\sum_f d_f(d_f-1)$ with $d_f=\binom{N}{f}$, plus one for the zero-frequency component, giving
\begin{equation}
    D_K = \binom{2N}{N} - 2^N + 1,
\end{equation}
where the Vandermonde identity $\sum_f \binom{N}{f}^2 = \binom{2N}{N}$ has been used. This matches the results from exact
diagonalization for every disorder realization. The fraction of nonvanishing matrix elements of $h$ in the energy eigenbasis
is $\binom{2N}{N}/4^N$, giving approximately $19.6\%$ at $N=8$ and $16.1\%$ at $N=12$. The hopping operator therefore sits
well below the naive Krylov bound, reflecting its block-diagonal structure in fermion number. When restricted to a fixed
fermion number sector, the operator saturates the Krylov bound even in the mass-deformed model.

\subsection{Lanczos Sequence}
\label{subsec:lanczos_coefficients_results}

Given the initial operator $\mathcal{O}$, the Lanczos algorithm produces a sequence of Lanczos coefficients $b_n$. In the
experiments below, the algorithm was run until the sequence terminates for the undeformed model and its deformations. Reaching large
iteration numbers requires high numerical precision, as errors accumulate across iterations. The computations here were performed
with between $256$ and $512$ decimal places of precision. Working at such precision is computationally expensive, which limits the maximum system
size $N$ that can be studied. To reduce the number of reorthogonalizations and lower the overall computational cost, the partial
reorthogonalization variant of the Lanczos algorithm was used. Appendix~\ref{app:lanczos_algorithms_comparison} reviews the differences
between full and partial reorthogonalization and compares their performance on $\mathcal{N}=2$ SYK.

Before studying how the Lanczos sequence changes with deformation strength, it is useful to first consider the undeformed model.
As discussed previously, the $\mathcal{H}_f^+$, $\mathcal{H}_f^-$, and $\mathcal{H}_f^z$
sectors are statistically independent and should be treated separately. The hopping operator does not
commute with the supercharges and so mixes these sectors, which motivates decomposing it into its projections within and
between sectors. For fermion number $f$, the relevant components are
\begin{equation}
    \label{eq:hopping_projected}
    h^{++} = (P_f^+)^\dagger h P_f^+,\quad h^{+-} = (P_f^-)^\dagger h P_f^+,\quad h^{+z} = (P_f^z)^\dagger h P_f^+,
\end{equation}
where $P_f^\pm$ and $P_f^z$ are the spectral projectors onto the $\mathcal{H}_f^\pm$ and $\mathcal{H}_f^z$ sectors.
The first operator maps $\mathcal{H}_f^+$ to itself, while the others map between different sectors. The latter two are
not Hermitian, but can be symmetrized by adding their Hermitian conjugates. For example,
\begin{equation}
    \label{eq:hopping_projected_sym}
    h^{+-,\text{sym}} = h^{+-} + h^{-+} = \begin{pmatrix}
0 & h^{-+} \\
h^{+-} & 0
    \end{pmatrix},
\end{equation}
and similarly for $h^{+z,\text{sym}}$, which are both traceless. Because $H$ annihilates the BPS block, the Liouvillian
acts one-sidedly on the $+z$ block:
\begin{equation}
    \mathcal{L}\,|E^+_a\rangle\langle z| = H|E^+_a\rangle\langle z| - |E^+_a\rangle\langle z|H = E^+_a\,|E^+_a\rangle\langle z|,
\end{equation}
so the Krylov problem for $h^{+z,\text{sym}}$ is unitarily equivalent to the spread complexity of a state evolving under $H^+ \oplus (-H^+)$.
The operator $h^{++}$ is not traceless in general, so its trace was subtracted before running the Lanczos algorithm in order to
eliminate the contribution from the one-point function, as discussed previously. Since the Hamiltonian commutes with the spectral projectors, the Liouvillian
preserves each subspace, ensuring the Lanczos algorithm remains within the appropriate sector. The remaining components
$h^{--}$ and $h^{-z}$ are statistically similar to $h^{++}$ and $h^{+z}$, respectively, and
the component $h^{zz}$ is trivial from the perspective of the Lanczos algorithm, as the Hamiltonian vanishes in the BPS sector.
For notational brevity, the symmetrized operators will be referred to simply as $h^{+-}$ and $h^{+z}$ in the discussion below.

\begin{figure}[bt]
    \centering
    \ifgeneratefigurerstatoperator
        \tikzsetnextfilename{new_figure_r_stat_operator}
        \input{figures/r_dist_zz_+-_+z_N12_f6_2.tex}
    \else
        \includegraphics{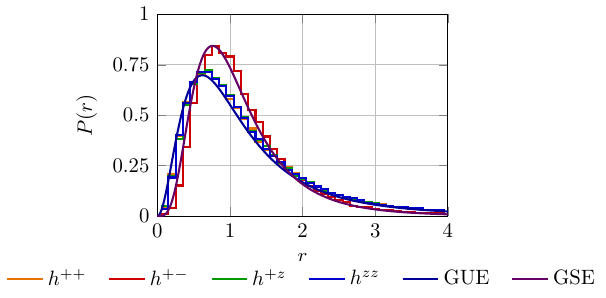}
    \fi
    \caption{Distribution of the gap ratio $r$ for the hopping operator projected to the $++$, $+-$, $+z$, and $zz$ sectors,
    for $N=12$ and $f=6$, computed using 500 disorder realizations. The GUE and GSE distributions are shown for reference.}
    \label{fig:r_stat_operator}
\end{figure}

Although $h^{zz}$ is invisible to the Lanczos algorithm, one can still examine its level statistics directly. Figure~\ref{fig:r_stat_operator}
shows the gap ratio distribution for $h^{zz}$, together with $h^{++}$, $h^{+-}$, and $h^{+z}$, for $N=12$ and $f=6$. All distributions except
$h^{+-}$ are consistent with GUE, while $h^{+-}$ is consistent with GSE. The difference can be attributed to charge conjugation symmetry,
which relates $\mathcal{H}_f^+$ and $\mathcal{H}_f^-$ and imposes an additional antiunitary symmetry on $h^{+-}$, enhancing the level
repulsion to $\beta = 4$. No such symmetry relates $\mathcal{H}_f^+$ and $\mathcal{H}_f^z$, leaving $h^{+z}$ in the GUE class. Similar
patterns are observed across the range of $N$ and $f$ values studied.

\begin{figure}[bt]
    \centering
    \setlength{\abovecaptionskip}{12pt}
    \ifgeneratefigurebnlongsequence
        \tikzsetnextfilename{new_figure_bn_long_sequence}
        \input{figures/bn_triptych.tex}
    \else
        \includegraphics{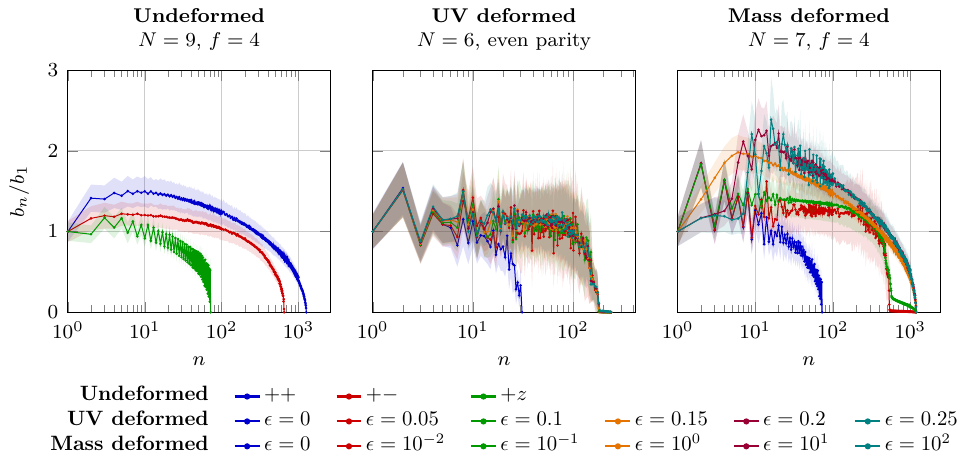}
    \fi
    \caption{The Lanczos coefficients $b_n$ for the hopping operator in the $\mathcal{N}=2$ SYK model
    and its deformations. Each curve corresponds to the average of 100 realizations of the Hamiltonian, with shaded regions indicating the
standard deviation. For the undeformed model, $D_K=1{,}261$ in the $++$ sector, $648$ in the $+-$ sector, and $72$ in the $+z$ sector.
For the UV-deformed model, $D_K = 33$ when the deformation is off and $241$ when on. For the mass-deformed model, $D_K = 75$ when the
deformation is off and $1{,}191$ when on. The values $q=3$ and $J=1$ were used throughout.}
    \label{fig:full_lanczos_sequence}
\end{figure}

The leftmost panel of Figure~\ref{fig:full_lanczos_sequence} shows the Lanczos coefficients for the $\mathcal{N}=2$ model
in the three sectors described above, for $N=9$ and $f=4$. The coefficients have been normalized by $b_1$, which sets the overall scale.
There are $d^{+}=36$ states in $\mathcal{H}_f^+$, $d^{-}=9$ in $\mathcal{H}_f^-$, and $81$ BPS states, with both the $\mathcal{H}_f^+$ and
$\mathcal{H}_f^-$ sectors belonging to the GUE class. The operator $h^{++}$ saturates its bound with $D_K = D_K^\text{max} = 1{,}261$, while
$h^{+-}$ and $h^{+z}$ do not saturate the bounds $D_K^\text{max} =1{,}981$ and $D_K^\text{max} =1{,}333$, respectively, as they are
unattainable for operators confined to a single off-diagonal block. The Krylov dimension is $2d^{+}d^{-}=2\cdot 36\cdot 9=648$ for $h^{+-}$ 
and $2d^{+}=72$ for $h^{+z}$, which saturate block off-diagonal bounds exactly. The sequences for $h^{++}$ and $h^{+-}$ are qualitatively similar,
while that for $h^{+z}$ exhibits more rapid oscillations. For $N=9$, the $\mathcal{H}_f^+$ sector at $f=3$ and the $\mathcal{H}_f^-$
sector at $f=6$ belong instead to the GSE class. This symmetry is broken by $h^{++}$ and $h^{+z}$, so no qualitative difference
is expected in their Lanczos sequences. For $h^{+-}$, which carries a time-reversal symmetry, a qualitative difference may be present.

The second panel of Figure~\ref{fig:full_lanczos_sequence} shows the complete Lanczos sequence under the UV deformation.
Since this deformation breaks the $U(1)$ symmetry, it is no longer possible to work in fixed fermion number sectors, but
fixed fermion parity sectors remain accessible. The even parity sector was used here. The plot displays the Lanczos sequences
for several values of $\epsilon$, up to the closing of the energy gap around $\epsilon \approx 0.25$. There is no significant
qualitative difference between the sequences across the different values of $\epsilon$. For $N=6$, all states belong to the
GSE class, though no qualitative difference is expected when the states are GOE instead, as occurs when $N\text{ mod }4 = 0,3$.

The third panel of Figure~\ref{fig:full_lanczos_sequence} shows the complete Lanczos sequence under the mass deformation,
with $N=7$ and $f=4$. The plot illustrates how the sequence evolves as $\epsilon$ is varied from the undeformed model through
the GUE regime to the Poisson regime, as shown in Figure~\ref{fig:r_stat_susy_ir}. At small but non-zero $\epsilon$, the sequence
grows erratically over the first few iterations before dropping sharply around $n \sim 500$, yet does not terminate until
$n = 1{,}191$, leaving a long flat tail of small coefficients. At $\epsilon = 1$, where the level statistics are GUE, the
ascent is smoother and more prolonged. At large $\epsilon$, where the statistics are Poisson, the characteristic ascent disappears
entirely and the sequence becomes noisy and erratic, as expected for a non-chaotic system.

\subsection{Krylov Complexity and Entropy}
\label{subsec:krylov_complexity_entropy_results}

\begin{figure}[bt]
    \centering
    \setlength{\abovecaptionskip}{12pt}
    \ifgeneratefigurefullkrylovcomplexity
        \tikzsetnextfilename{new_figure_full_krylov_complexity}
        \input{figures/krylov_complexity_triptych.tex}
    \else
        \includegraphics{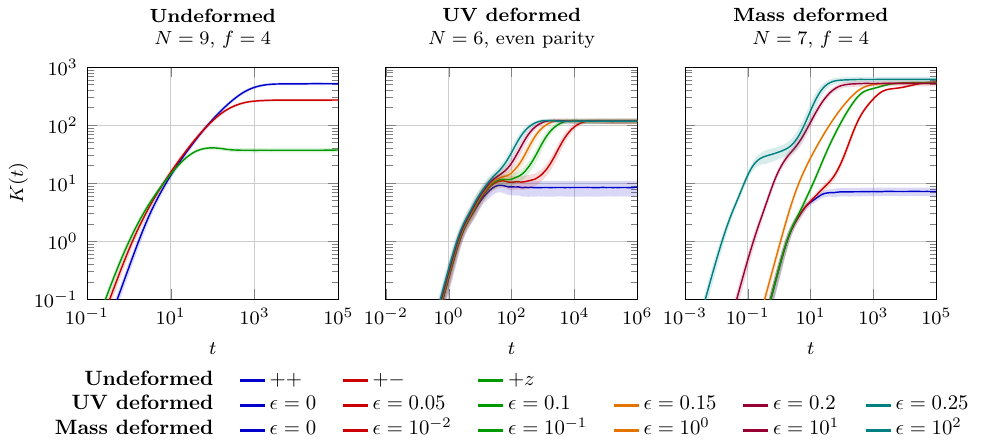}
    \fi
    \caption{Krylov complexity $K(t)$ for the hopping operator in the $\mathcal{N}=2$ SYK model and its deformations.
Each curve corresponds to the average of 100 realizations of the Hamiltonian, with shaded regions indicating the
standard deviation.}
    \label{fig:full_krylov_complexity}
\end{figure}

Once the Lanczos coefficients have been computed, the dynamics of the operator in the Krylov lattice are fully determined.
The Krylov wavefunctions $\varphi_n(t)$ were obtained by integrating Equation~\eqref{eq:hopping_equation} numerically and
substituted into Equation~\eqref{eq:krylov_complexity} to give the Krylov complexity $K(t)$, which corresponds to the
average position of the operator in the Krylov lattice. This procedure was repeated for each disorder realization and
the results averaged at the end. This quenched disorder averaging is standard in studies of Krylov complexity. Averaging the
sequences before computing complexity can prevent the wavepacket from randomizing efficiently before reaching the
end of the Krylov lattice, potentially introducing large unphysical oscillations in $K(t)$~\cite{Rabinovici_2021}. In the cases
considered here, the two orderings were found to agree.

Figure~\ref{fig:full_krylov_complexity} shows Krylov complexity for the hopping operator in the $\mathcal{N}=2$ SYK model
and its deformations. In the undeformed model, shown in the leftmost panel, the complexity grows and saturates at a well-defined plateau,
with the $+z$ sector reaching a noticeably smaller saturation value than the $++$ and $+-$ sectors. The middle panel shows the UV deformation,
where turning on $\epsilon$ raises the saturation value significantly, though it does not change appreciably as $\epsilon$ is increased further.
The rightmost panel shows the mass deformation, where the saturation complexity rises as the deformation is turned on. In all cases, complexity grew initially quadratically
then linearly, consistent with behavior observed in systems with different symmetry classes \cite{Caputa_2021, Guo_2022}.

\begin{figure}[bt]
    \centering
    \setlength{\abovecaptionskip}{12pt}
    \ifgeneratefiguresaturationkrylovcomplexity
        \tikzsetnextfilename{new_figure_saturation_krylov_complexity}
        \input{figures/ksat_triptych.tex}
    \else
        \includegraphics{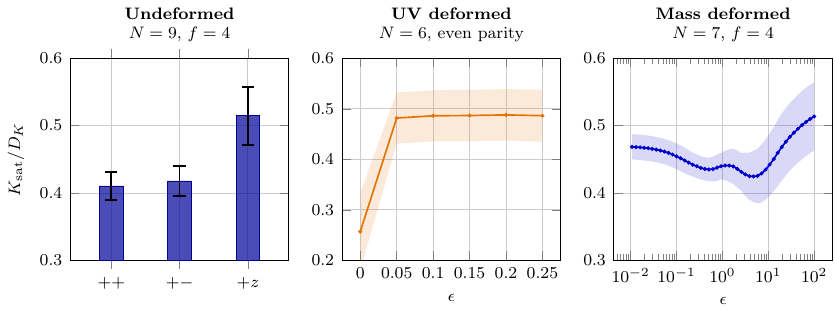}
    \fi
    \caption{The saturation value of Krylov complexity $K_\text{sat}$ for the hopping operator in the $\mathcal{N}=2$ SYK model and its deformations.
The saturation complexity is reported as a fraction of the Krylov dimension $D_K$ for each model. For lower values of $\epsilon$, $K_\text{sat}$ for
the mass-deformed model hovers around $0.48$. The complexity was computed from 100 realizations
of the Hamiltonian and corresponding Lanczos sequences.}
    \label{fig:saturation_krylov_complexity}
\end{figure}

Figure~\ref{fig:saturation_krylov_complexity} shows the saturation Krylov complexity $K_\text{sat}$ as a fraction of the Krylov dimension
$D_K$ for the various models.  In the undeformed model, the $+z$ sector achieves $K_\text{sat}/D_K  = 0.51 \pm 0.04$, which is close to the spread value of
one-half for a state delocalized over a chaotic spectrum. The $++$ and $+-$ sectors give roughly $0.41$. Before the UV deformation is turned on,
the saturation complexity in the full $N=6$ parity-even sector is suppressed to $0.26 \pm 0.08$, reflecting mixing of statistically independent
sectors across different fermion numbers. Once the deformation
is turned on, it rises to $0.48 \pm 0.05$. In the mass-deformed case with $N=7$ and $f=4$, the pre-deformation value is $0.10 \pm 0.02$,
suppressed by mixing of statistically independent sectors within the same fermion number. Turning on the deformation raises it to around
$0.47$ at $\epsilon = 10^{-2}$. It then decreases  to a shallow minimum at $\epsilon \sim 1$ before rising again.

\begin{figure}[bt]
    \centering
    \ifgeneratefigurealphaVKcombined
        \tikzsetnextfilename{new_figure_alpha_vK_combined}
        \input{figures/alpha_vK_combined.tex}
    \else
        \includegraphics{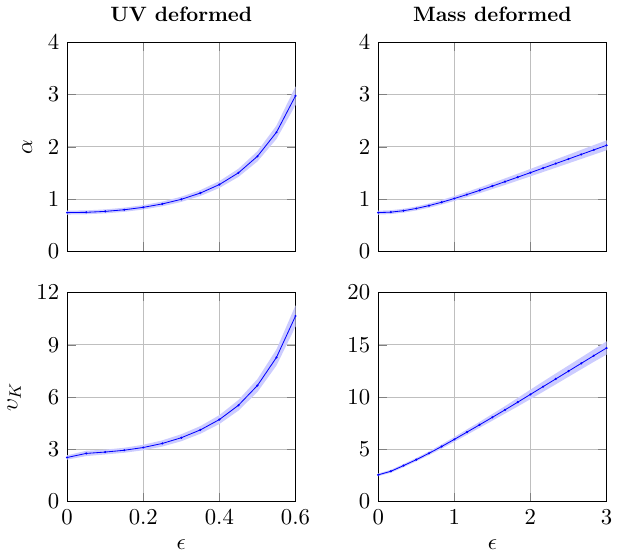}
    \fi
    \caption{Comparison of Krylov complexity growth parameters of the hopping operator for
    $\mathcal{N} = 2$ SYK as the deformation parameter $\epsilon$ varies. Prior to $t \sim t_*$, the
    growth is given by a power law $K(t) \sim \alpha^2 t^2$. After $t_*$, complexity grows linearly,
    $K(t) \sim \upsilon_Kt$. Results are reported separately for UV and mass deformations.}
    \label{fig:krylov_growth_rate_parameters_epsilon_sweep}
\end{figure}

It is also possible to compute the Lanczos coefficients and Krylov complexity in the full model without restricting to a
fixed fermion number or parity sector, which serves as a check on the expected growth profile. For both deformations, the
complexity grew initially quadratically before transitioning to linear growth, corresponding in the notation of
Table~\ref{tab:krylov-behaviors} to the Krylov exponent $\delta$ transitioning from one-half to zero. The transition was
identified by computing $\delta$ over time from smoothed estimates of $K(t)$ and its derivatives, and the growth coefficients
in each phase were extracted following Appendix~\ref{app:krylov_complexity_parameters_estimation}.
Figure~\ref{fig:krylov_growth_rate_parameters_epsilon_sweep} shows the quadratic coefficient $\alpha$ and the linear
ballistic velocity $\upsilon_K$ as functions of $\epsilon$. For the UV deformation, both $\alpha$ 
and $\upsilon_K$ grow linearly with $\gamma$. Under the mass deformation, both $\alpha$ and $\upsilon_K$ grow roughly
linearly with $\epsilon$.

From the Krylov wavefunctions, the Krylov entropy $S(t)$ was also computed using Equation \eqref{eq:krylov_entropy}. At early times,
the wavefunction has spread out slightly from its initial value $\varphi_n(0) = \delta_{n0}$, and one can treat $\varphi_0(t)$
and $\varphi_1(t)$ as non-zero. Equation \eqref{eq:hopping_equation} then reduces to the system
\begin{equation}
    \dot{\varphi}_0(t) = -b_1 \varphi_1(t),\quad \dot{\varphi}_1(t) = b_1 \varphi_0(t).
\end{equation}
This gives $\varphi_0(t) = \cos(b_1 t)$, which leads to $|\varphi_0(t)|^2 \approx 1 - b_1^2 t^2$ and
$|\varphi_1(t)|^2 \approx b_1^2 t^2$. The Krylov entropy is given by
\begin{equation}
    \label{eq:krylov_entropy_early_times}
    S(t) \approx b_1^2t^2\left(1 - 2\log(b_1t)\right),
\end{equation}
which is quadratic with a logarithmic correction. This quasi-quadratic behavior was observed in the data at very small time scales
in the full models. The observed growth scales like $S(t) \sim t^{1.7}$ across all deformations, consistent with sub-quadratic growth.
Fitting Equation \eqref{eq:krylov_entropy_early_times} to the entropy at early times also gives values of $b_1$ consistent with those
shown in Figure \ref{fig:krylov_growth_rate_parameters_epsilon_sweep}.

\begin{figure}[bt]
    \centering
    \setlength{\abovecaptionskip}{12pt}
    \ifgeneratefigurefullkryloventropy
        \tikzsetnextfilename{new_figure_full_krylov_entropy}
        \input{figures/krylov_entropy_triptych.tex}
    \else
        \includegraphics{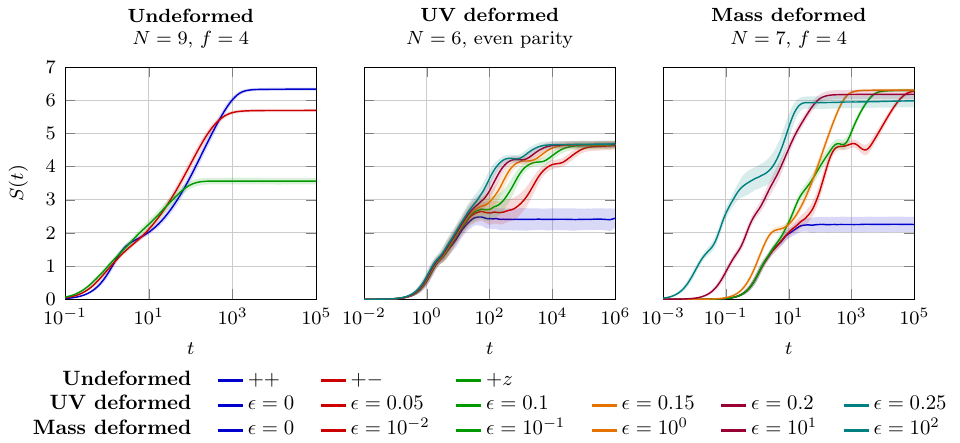}
    \fi
    \caption{Krylov entropy $S(t)$ for the hopping operator in the $\mathcal{N}=2$ SYK model and its deformations.
Each curve corresponds to the average of 100 realizations of the Hamiltonian, with shaded regions indicating the standard deviation.}
    \label{fig:full_krylov_entropy}
\end{figure}

\begin{figure}[bt]
    \centering
    \setlength{\abovecaptionskip}{12pt}
    \ifgeneratefiguresaturationkryloventropy
        \tikzsetnextfilename{new_figure_saturation_krylov_entropy}
        \input{figures/ssat_triptych.tex}
    \else
        \includegraphics{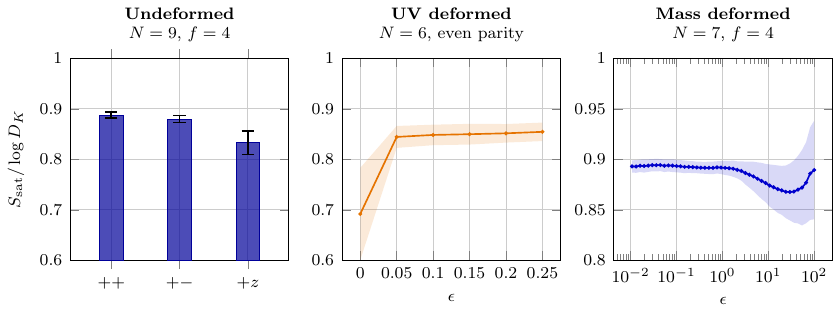}
    \fi
    \caption{The saturation value of Krylov entropy $S_\text{sat}$ for the hopping operator in the $\mathcal{N}=2$ SYK model and its deformations.
    The saturation entropy is reported as a fraction of the log of the Krylov dimension $D_K$ for each model. The entropy was computed
    from 100 realizations of the Hamiltonian and corresponding Lanczos sequences.}
    \label{fig:saturation_krylov_entropy}
\end{figure}

Figure~\ref{fig:full_krylov_entropy} shows the full Krylov entropy profile for the hopping operator in the $\mathcal{N}=2$ SYK model and its
deformations, and Figure~\ref{fig:saturation_krylov_entropy} shows the saturation entropy $S_\text{sat}$ for each model as a fraction of $\log D_K$.
The latter is the expected scaling when the late-time wavefunction spreads uniformly across the Krylov lattice. In the undeformed model, the entropy
is lowest in the $+z$ sector. As with the saturation complexity, $S_\text{sat}$ jumps when the UV deformation is turned on, remaining consistent with
$S_\text{sat} \sim \log K_\text{sat}$. Under the mass deformation, the entropy stays roughly constant until $\epsilon$ is of order one, after which
it decreases before rising again. At small $\epsilon$, the entropy is consistent with logarithmic scaling, but deviates from it at larger $\epsilon$,
reflecting increased randomization of the wavefunction in the Krylov lattice.

It should be emphasized that the results presented above were computed at different system sizes, chosen so that the relevant Krylov dimensions remain
tractable at the precision required for convergence. At the sizes accessible here, the Lanczos ascent is short and the exponential growth regime
of the large-$N$ chaotic limit is not reached. The early-time quadratic growth of $K(t)$ is universal, so the quadratic to linear transition is
generic to bounded finite systems and does not, by itself, separate chaotic from integrable dynamics. The chaotic-versus-integrable conclusions discussed
in this work rest primarily on the level statistics and on the late-time saturation complexity.

\section{Discussion}
\label{sec:conclusion}

This paper explored the effect of supersymmetry breaking on Krylov complexity in the $\mathcal{N}=2$ SYK model using two
types of deformations of the Hamiltonian. The first deformation was an irrelevant deformation which broke $\mathcal{N}=2$
supersymmetry down to $\mathcal{N}=1$, but preserved a remnant chiral symmetry of the self-adjoint supercharge. The deformation
continuously deformed the anticommutation relations between fermions, effectively rotating them relative to canonical fermions.
In a holographic context, the deformation preserved the gravitational sector which occurs in the infrared. The second deformation
was a mass deformation which broke $\mathcal{N}=2$ down directly to $\mathcal{N}=0$ through the introduction of a free Hamiltonian,
but preserved the global $U(1)$ symmetry. The deformation augmented the gravitational picture by modifying the low-energy dynamics.

The effect of the deformations was first studied through the level statistics and random matrix properties of each theory. The
undeformed $\mathcal{N}=2$ SYK model has a large degeneracy of zero-energy BPS states. In the non-BPS sector, states are divided
according to whether they are annihilated by $Q$ or $Q^\dagger$, and the symmetry class within this sector is GUE, GSE, or GOE
depending on $N$ and $f$, with a two- to four-fold degeneracy. The UV deformation lifts the BPS degeneracy and raises the ground
state energy while keeping it positive. The gap between the BPS and non-BPS states closes continuously with increasing $\epsilon$,
vanishing around $\epsilon \approx 0.25$, and the deformed spectrum is entirely non-BPS with GOE or GSE statistics. The mass
deformation also lifts the BPS degeneracy, but drives the ground state energy negative even for parametrically small deformation
strengths, signaling a complete breaking of supersymmetry. As $\epsilon$ increases, the level statistics transition first to a GUE
regime and then to Poisson at large deformation strengths, with the degeneracy structure completely broken, resembling the
$\mathcal{N}=0$ SYK model.

Following this preliminary analysis, a Krylov subspace analysis was conducted on both deformations, which explored the
Krylov dimension, Lanczos sequence, and Krylov complexity and entropy. The results are summarized as follows.

\begin{enumerate}
    \item \textbf{Krylov dimension}. The UV deformation reduced the energy degeneracy and increased $D_K^\text{max}$,
while the mass deformation reduced the degeneracy to one, raising the bound even further. The hopping operator saturated $D_K^\text{max}$
in the UV-deformed model, but fell below the bound in both the undeformed and mass-deformed cases.
    \item \textbf{Lanczos sequence}. Under the UV deformation, the Lanczos coefficients showed qualitatively similar growth
across all deformations up to the closing of the energy gap, where the statistics are GOE or GSE. Under the mass
deformation, the ascent is smoother and more prolonged at $\epsilon = 1$ where the statistics are GUE, and the ascent
disappears as $\epsilon$ increases into the Poisson regime. In the full models, the Lanczos coefficients exhibited
initial sublinear growth $b_n \sim \alpha n^\delta$ with $\delta = 0.5$ across both deformations, with the growth rate $\alpha$
increasing with $\epsilon$.

    \item \textbf{Krylov complexity and entropy}. In the undeformed model, the saturation Krylov complexity is highest in
the sector mixing BPS and non-BPS states. The UV deformation raises the saturation complexity to roughly half the Krylov dimension,
consistent with a wavefunction spread uniformly across the Krylov lattice, and the saturation entropy increases accordingly.
Under the mass deformation, the saturation complexity decreases, exhibiting a local minimum near $\epsilon \sim 1$. In the full models, Krylov complexity grew initially
as $t^2$ before transitioning to linear growth, with both rates increasing under either deformation.
\end{enumerate}

To summarize, the impact of supersymmetry breaking on Krylov complexity of the $\mathcal{N}=2$ SYK model depends on the
specific mechanism of symmetry breaking. 
In both cases the saturation complexity rises as
soon as the deformation is turned on. Under the UV deformation, it then plateaus at $K_\text{sat}/D_K \approx 0.5$ for all deformation strengths.
Under the mass deformation, this initial rise gives way to a decrease before rising again at larger $\epsilon$.

In the undeformed model, the $+z$ sector, which mixes $\mathcal{H}_f^+$ and BPS states, achieves
the highest saturation complexity ratio $K_\text{sat}/D_K \approx 0.51$ among the three sectors, yet has the lowest saturation
entropy ratio $S_\text{sat}/\log D_K$. This is because the two quantities measure different aspects of the Krylov wavefunction.
The complexity measures its mean position, while the entropy measures how broadly it is spread. The BPS states, whose Hamiltonian
vanishes identically, do not propagate in the usual sense and may restrict the effective spreading of the wavefunction across
the Krylov lattice, leading to a higher mean position but a more concentrated distribution. This suggests that the interplay between
BPS and non-BPS states may leave an imprint on the structure of the Krylov wavefunction in the undeformed model.

There are several ways this work can be extended in the future.

\begin{enumerate}
    \item \textbf{Finite temperature}. This work employed the Lanczos algorithm using an inner product that corresponds to
infinite temperature. It would be interesting to repeat the analysis using a finite-temperature version of the Lanczos algorithm. Working at
finite temperature naively breaks supersymmetry, but it is still possible to study the Lanczos sequence and Krylov
complexity at finite temperature, which could shed light on the strong coupling regime of the theory, $\beta J \gg 1$.
    \item \textbf{Other complexity measures}. This work focused on Krylov complexity, but there are other definitions of complexity,
including out-of-time-order correlation functions which are characterized by the Lyapunov exponent. It would be interesting
to see how the Lyapunov exponent responds to supersymmetry and its breaking in $\mathcal{N}=2$ SYK. For example, without supersymmetry,
the mass deformation causes the Lyapunov exponent to go to zero at low temperatures, while the Krylov exponent continues to
be near-maximal \cite{Chapman_2024}. It would be interesting to see if this behavior persists in the supersymmetric case. Other
complexity measures, including circuit complexity and Nielsen complexity, can also be explored \cite{Baiguera_2025, Balasubramanian2021, Craps2024}.
    \item \textbf{Symmetry-resolved Krylov complexity}. Another way to study the effect of symmetries on Krylov complexity is to
analyze symmetry-resolved Krylov complexity, which captures the impact each symmetry sector has on complexity \cite{Caputa_2025, Caputa_2025_2}.
Generally, complexity at early times equals the average over symmetry sectors, but at late times the interplay between sectors
becomes important. Symmetry-resolved Krylov complexity of $\mathcal{N}=2$ SYK can be studied across sectors of fixed
$U(1)$ or $U(1)_R$ charge. Breaking supersymmetry usually also breaks these symmetries, but it would be interesting to study the
contribution of each sector to the overall complexity when supersymmetry is intact.
    \item \textbf{Holographic dual}. This paper studied the quantum mechanical notion of complexity, but the most
interesting area of future research would be to explore the holographic analogue of Krylov complexity in the dual
gravitational system \cite{Jian_2021, Rabinovici_2023, Ambrosini_2025, Ambrosini_2025_2, Caputa_2024, Heller2025a, Heller2025b,
Aguilar-Gutierrez2025b, Aguilar-Gutierrez2025c, Aguilar-Gutierrez2025d, Aguilar-Gutierrez2024}. In a two-sided system consisting
of two copies of the SYK model, the holographic dual to Krylov complexity has been interpreted as the size of the wormhole
connecting the boundaries. The supersymmetric version of this picture involves super-JT gravity and the size of supersymmetric
wormholes \cite{Aguilar-Gutierrez2025}. It would be interesting to interpret the deformations of supersymmetric SYK treated
here in terms of the growth of supersymmetric wormholes in the bulk.
\end{enumerate}

\noindent \textbf{Acknowledgments.} The authors would like to thank Debarghya Chakraborty, Johanna Erdmenger, Dami\'an Galante,
Christian Northe, and Martin Sasieta for useful discussions. The authors also thank the organizers of the ``Quantum Gravity,
Holography, and Quantum Information'' conference at the International Institute of Physics, Rio Grande do Norte, Brazil,
where some of the results were presented. The authors are also grateful to the anonymous referee, whose thoughtful and
constructive feedback led to a substantially improved version of this manuscript. The work of DV is partially supported
by the STFC Consolidated Grant ST/X00063X/1 ``Amplitudes, Strings, \& Duality.'' This research utilized Queen Mary's Apocrita
HPC facility, supported by QMUL Research-IT (\url{http://doi.org/10.5281/zenodo.438045}).  \\

\noindent \textbf{Data access statement.} The code and data used in this study are publicly available in the ancillary files at \url{https://arxiv.org/abs/2511.20769}.

\appendix

\clearpage
\section{Comparison of Different Lanczos Algorithms}
\label{app:lanczos_algorithms_comparison}

There are two common versions of the Lanczos algorithm used in Krylov subspace methods, the full orthogonalization (FO)
and the partial reorthogonalization (PRO) algorithms. This appendix summarizes the two algorithms and compares their
performance in the computation of the Lanczos coefficients for the $\mathcal{N}=2$ SYK model.

The full orthogonalization algorithm maintains orthogonality of the Krylov basis at each iteration by explicitly
orthogonalizing the new basis element against all previously computed elements. The new Krylov basis element is
given by
\begin{align}
    |\mathcal{A}_n) &= \mathcal{L}|\mathcal{O}_{n-1}), \\
    |\mathcal{A}_n) &\leftarrow |\mathcal{A}_n) - \sum_{i=0}^{n-1} |\mathcal{O}_i)(\mathcal{O}_i|\mathcal{A}_n). \label{eq:fo_reorthogonalization}
\end{align}
This update step replaces step two of the basic Lanczos algorithm outlined in Section \ref{subsec:krylov_space}, but
the rest of the algorithm remains unchanged. In practice, the orthogonalization of Equation \eqref{eq:fo_reorthogonalization} is
performed $n_\text{O} > 1$ times to ensure numerical stability. The FO algorithm is more accurate and numerically stable than the
basic Lanczos algorithm, but can be computationally expensive, especially for large Krylov subspaces.

To reduce the computational cost of computing the Krylov basis, the partial reorthogonalization algorithm reorthogonalizes
the new basis element only when the level of overlap with previous basis elements exceeds a certain threshold. To achieve this,
PRO estimates the overlap between the $n$th Krylov basis element and the $k$th element using the recurrence relation \cite{Rabinovici_2021}
\begin{equation}
    W_{kn} \equiv (\mathcal{O}_k|\mathcal{O}_n) = \frac{1}{b_n}\left(b_{k+1}W_{k+1,n-1}^* + b_k W_{k-1,n-1}^* -
    b_{n-1}W_{k,n-2}\right),\quad k\geq 0, n \geq 2,
\end{equation}
where $W_{-1,n} \equiv 0$. In this equation, the $n$th column of $W$ is computed using the previous two columns, requiring only two columns
of $W$ to be stored at any time. Because the recurrence relation does not determine the overlap with the previous
element $W_{n-1,n}$, the element $|\mathcal{A}_n)$ is explicitly orthogonalized against $|\mathcal{O}_{n-1})$ at each
iteration. The relation also does not determine $W_{nn}$, but it is set to unity after normalizing $|\mathcal{A}_n)$.

If $W_{kn}$ exceeds a predetermined threshold $\epsilon_\text{P}$ for any $k \leq n-2$, then $|\mathcal{A}_n)$ and
$|\mathcal{A}_{n-1})$ are reorthogonalized against all previous elements using Equation \eqref{eq:fo_reorthogonalization}.
As with the FO algorithm, this orthogonalization step is performed $n_\text{O}$ times. After reorthogonalization,
the overlap matrix elements are set to
\begin{equation}
    W_{ik} = \delta_{ik} + (1-\delta_{ik})\epsilon_\text{M},\quad \forall \, i \leq k, k \in \{n-1,n\}.
\end{equation}
The parameter $\epsilon_\text{M}$ is the machine precision of the floating point representation used in
the computation, which sets the overlap between different basis elements after reorthogonalization. The advantage of the
PRO algorithm is that it reduces the number of reorthogonalizations required while maintaining orthogonality of the Krylov basis
to a desired level of accuracy. In experiments on the UV and mass-deformed models, PRO required an order of magnitude fewer
reorthogonalizations than the FO algorithm after $200$ iterations. In practice, PRO is less accurate than FO, but significantly
more efficient, allowing larger Krylov subspaces to be computed. On a MacBook Pro with M1 Pro chip and $32$\,GB RAM,
the PRO algorithm ran roughly $14$--$23\%$ faster than the FO algorithm for the UV-deformed model, and around $22$--$55\%$ faster
for the mass-deformed model. While this is a modest speedup, it grows significantly as the number of iterations increases.

\begin{figure}[bt]
    \centering
    \includegraphics[keepaspectratio]{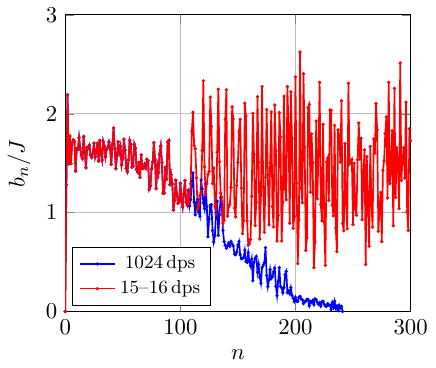}
    \caption{Comparison of the Lanczos coefficients $b_n$ computed using the FO algorithm using two different levels of numerical
    precision for the UV-deformed $\mathcal{N}=2$ SYK model with $\epsilon=0.6$. The low-precision computation used the IEEE 754
    double-precision floating-point format, which utilizes between $15$ and $16$ decimal places, or dps. The high-precision computation
    used $1024$\,dps. The values $N = 6$, $q = 3$, and $J = 1$ were used throughout.}
    \label{fig:lanczos_different_precisions}
\end{figure}

Regardless of the Lanczos algorithm used, computing the Lanczos coefficients often requires high numerical precision to
maintain stability. Insufficient precision can cause quantities such as $a_n \equiv (\mathcal{O}_n|\mathcal{L}|\mathcal{O}_n)$,
which should vanish for Hermitian operators $\mathcal{O}_n$, to drift away from zero. This drift produces inaccurate Lanczos
coefficients, particularly after many iterations. Low precision can also lead to the total number of iterations exceeding
the theoretical maximum Krylov dimension reported in Section~\ref{subsec:krylov_dimension_results}. This effect is illustrated in
Figure \ref{fig:lanczos_different_precisions}, which compares the Lanczos coefficients computed using the FO algorithm
with two different levels of numerical precision for the UV-deformed $\mathcal{N}=2$ SYK model with $\epsilon=0.6$.
The high-precision computation terminates within the Krylov bound of $241$, while the low-precision computation exceeds the bound.
This example highlights the importance of using sufficient numerical precision when computing the Lanczos coefficients to ensure
accurate results. Utilizing high precision also increases the runtime of the algorithms, effectively limiting
the system sizes that can be simulated.

\vspace{2em}

\section{Estimation of Krylov Complexity Parameters}
\label{app:krylov_complexity_parameters_estimation}

To analyze the growth of Krylov complexity, the various parameters characterizing the growth need to be estimated.
At finite $N$, Krylov complexity follows power-law growth given by $K(t) \sim (\alpha t)^\frac{1}{1-\delta}$. One way to
estimate the parameters is to use local estimates of the complexity and its derivatives. Complexity and its first two
derivatives satisfy the following relations:
\begin{align}
    K(t) &= (\alpha t)^\frac{1}{1-\delta} + K_0, \\
    K'(t) &= \frac{(\alpha t)^{\frac{1}{1-\delta}}}{(1-\delta) t}, \\
    K''(t) &= \frac{\delta (\alpha t)^{\frac{1}{1-\delta}}}{(\delta - 1)^2 t^2}.
\end{align}
This system of equations can be solved to give the parameters in terms of $K(t)$ and its derivatives:
\begin{align}
    \alpha(t) &= \left( \frac{{K'}^{2}\, t^{-\tfrac{K'' t}{K'}}}{K' + K'' t} \right)^{\tfrac{K'}{K' + K'' t}}, \label{eq:alpha_estimate} \\
    \delta(t) &= \frac{K'' t}{K' + K'' t}, \label{eq:delta_estimate} \\
    K_0(t) &= \frac{K K' - (K')^{2} t + K K'' t}{K' + K'' t}.
\end{align}

\begin{figure}[bt]
    \centering
    \ifgeneratefigurekrylovexponentcombined
        \tikzsetnextfilename{new_figure_krylov_exponent_combined}
        \input{figures/krylov_exponent_combined.tex}
    \else
        \includegraphics{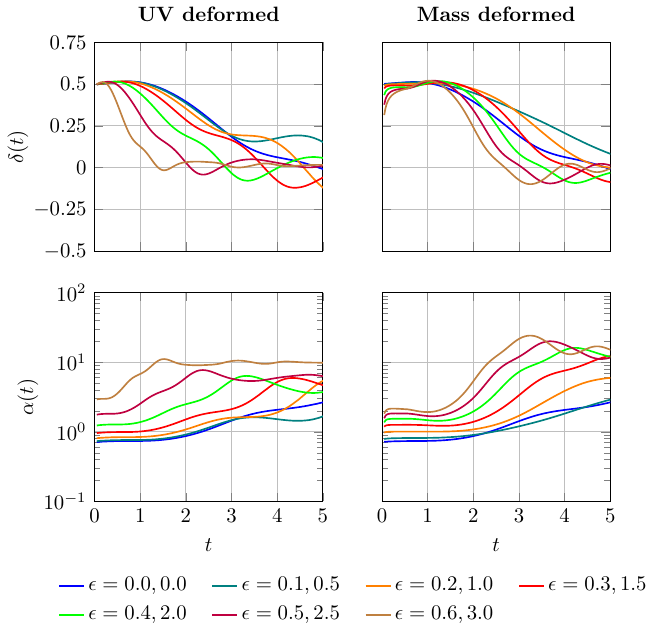}
    \fi
    \caption{Krylov exponent $\delta(t)$ and coefficient $\alpha(t)$ of $\mathcal{N}=2$ SYK for the hopping operator $h_{N-1,N}$, estimated
    using the smoothed complexity $K(t)$ and its smoothed first and second derivatives. The effects of UV and mass
    deformations on the exponent are shown as the deformation strength $\epsilon$ is varied. The first value of $\epsilon$
    in the legend corresponds to the UV deformation, and the second value to the mass deformation.}
    \label{fig:krylov_delta_N2_SUSY_SYK_N_12}
\end{figure}

In practice, $K(t)$ and its derivatives need to be smoothed to obtain reliable estimates of the parameters. The first row of Figure
\ref{fig:krylov_delta_N2_SUSY_SYK_N_12} shows estimates of $\delta(t)$ given by Equation \eqref{eq:delta_estimate} for
the hopping operators and the two deformations. Smoothing was performed using a Savitzky-Golay filter, which uses least squares to
fit the data over a moving window with low-degree polynomials. The moving window was set to $0.1$ in units of $1/J$ for the UV
deformation, and one for the mass deformation. Cubic polynomials were used for both deformations. In all instances,
the exponent starts around a half, then relaxes to values around zero, which corresponds to Krylov complexity transitioning
between quadratic and linear growth. The transition is not abrupt, but occurs over the course of one to two units,
depending on the deformation. Because the transition is gradual, it is impractical to estimate a single scrambling time.
In general, the irrelevant deformation advances the start of the linear growth phase. The mass deformation delays the
start of the linear growth phase for small values of $\epsilon$, but begins moving it up for
larger values.

The second row of Figure \ref{fig:krylov_delta_N2_SUSY_SYK_N_12} shows the estimates of $\alpha(t)$ given by Equation \eqref{eq:alpha_estimate}.
The values of $\alpha$ at early times correspond to the growth rate in the quadratic phase, while the values at late
times correspond to the growth rate in the linear phase, denoted $\upsilon_K$ in the main text. Because these direct
estimates can be noisy, an alternative method to estimate $\alpha$ and $\upsilon_K$ was employed based on $\delta(t)$. In
this method, the times at which $\delta(t)$ crosses the thresholds of $0.5$ and zero were identified. In practice, the
thresholds were adjusted to account for noise in the estimates. The two crossing times define the intervals over which
quadratic and linear fits to the complexity data were performed. To improve robustness, multiple fits were carried out
by varying the interval endpoints, and the final estimates were taken as the averages over all fits. The results are
summarized in Figure \ref{fig:krylov_growth_rate_parameters_epsilon_sweep} of the main text.

\clearpage
\bibliographystyle{unsrtnat}
\bibliography{references_krylov_complexity_susy_v2}

@article{Parker_2019,
    author = "Parker, Daniel E. and Cao, Xiangyu and Avdoshkin, Alexander and Scaffidi, Thomas and Altman, Ehud",
    title = "{A Universal Operator Growth Hypothesis}",
    eprint = "1812.08657",
    archivePrefix = "arXiv",
    primaryClass = "cond-mat.stat-mech",
    doi = "10.1103/PhysRevX.9.041017",
    journal = "Phys. Rev. X",
    volume = "9",
    number = "4",
    pages = "041017",
    year = "2019",
   note={\href{https://arxiv.org/abs/1812.08657}{\texttt{[1812.08657]}}}
}

@article{MSS_2016,
    author = "Maldacena, Juan and Shenker, Stephen H. and Stanford, Douglas",
    title = "{A bound on chaos}",
    eprint = "1503.01409",
    archivePrefix = "arXiv",
    primaryClass = "hep-th",
    doi = "10.1007/JHEP08(2016)106",
    journal = "JHEP",
    volume = "08",
    pages = "106",
    year = "2016",
    note={\href{https://arxiv.org/abs/1503.01409}{\texttt{[1503.01409]}}}
}

@article{Nandy_2024,
   author = "Nandy, Pratik and Matsoukas-Roubeas, Apollonas S. and Mart{\'\i}nez-Azcona, Pablo and Dymarsky, Anatoly and del Campo, Adolfo",
    title = "{Quantum dynamics in Krylov space: Methods and applications}",
    eprint = "2405.09628",
    archivePrefix = "arXiv",
    primaryClass = "quant-ph",
    reportNumber = "RIKEN-iTHEMS-Report-24",
    doi = "10.1016/j.physrep.2025.05.001",
    journal = "Phys. Rept.",
    volume = "1125-1128",
    pages = "1--82",
    year = "2025",
   note={\href{https://arxiv.org/abs/2405.09628}{\texttt{[2405.09628]}}}
}

@article{Rabinovici_2023,
    author = "Rabinovici, E. and S{\'a}nchez-Garrido, A. and Shir, R. and Sonner, J.",
    title = "{A bulk manifestation of Krylov complexity}",
    eprint = "2305.04355",
    archivePrefix = "arXiv",
    primaryClass = "hep-th",
    doi = "10.1007/JHEP08(2023)213",
    journal = "JHEP",
    volume = "08",
    pages = "213",
    year = "2023",
   note={\href{https://arxiv.org/abs/2305.04355}{\texttt{[2305.04355]}}}
}

@article{Ambrosini_2025,
    author = "Ambrosini, Marco and Rabinovici, Eliezer and S{\'a}nchez-Garrido, Adri{\'a}n and Shir, Ruth and Sonner, Julian",
    title = "{Operator K-complexity in DSSYK: Krylov complexity equals bulk length}",
    eprint = "2412.15318",
    archivePrefix = "arXiv",
    primaryClass = "hep-th",
    reportNumber = "CERN-TH-2025-040",
    doi = "10.1007/JHEP08(2025)059",
    journal = "JHEP",
    volume = "08",
    pages = "059",
    year = "2025",
   note={\href{https://arxiv.org/abs/2412.15318}{\texttt{[2412.15318]}}}
}

@article{Sachdev_1993,
   title={{Gapless Spin-fluid Ground State in a Random Quantum Heisenberg Magnet}},
  volume={70},
   ISSN={0031-9007},
   url={http://dx.doi.org/10.1103/PhysRevLett.70.3339},
   DOI={10.1103/physrevlett.70.3339},
   number={21},
   journal={Physical Review Letters},
   publisher={American Physical Society (APS)},
   author={Sachdev, Subir and Ye, Jinwu},
   year={1993},
   month=May, pages={3339–3342},
   note={\href{https://arxiv.org/abs/cond-mat/9212030}{\texttt{[cond-mat/9212030]}}}
}

@article{Polchinski_2016,
    author = "Polchinski, Joseph and Rosenhaus, Vladimir",
    title = "{The Spectrum in the Sachdev-Ye-Kitaev Model}",
    eprint = "1601.06768",
    archivePrefix = "arXiv",
    primaryClass = "hep-th",
    doi = "10.1007/JHEP04(2016)001",
    journal = "JHEP",
    volume = "04",
    pages = "001",
    year = "2016",
   note={\href{https://arxiv.org/abs/1601.06768}{\texttt{[1601.06768]}}}
}

@article{Maldacena_2016,
    author = "Maldacena, Juan and Stanford, Douglas",
    title = "{Remarks on the Sachdev-Ye-Kitaev model}",
    eprint = "1604.07818",
    archivePrefix = "arXiv",
    primaryClass = "hep-th",
    doi = "10.1103/PhysRevD.94.106002",
    journal = "Phys. Rev. D",
    volume = "94",
    number = "10",
    pages = "106002",
    year = "2016",
   note={\href{https://arxiv.org/abs/1604.07818}{\texttt{[1604.07818]}}}
}

@misc{kitaev2015talks,
      author         = "Kitaev, A",
      title          = "{A Simple Model of Quantum Holography,
       Talks at KITP, April 7, 2015 and May 27, 2015,
      http://online.kitp.ucsb.edu/online/entangled15/kitaev}"
}

@article{Cotler_2017,
    author = "Cotler, Jordan S. and Gur-Ari, Guy and Hanada, Masanori and Polchinski, Joseph and Saad, Phil and Shenker, Stephen H. and Stanford, Douglas and Streicher, Alexandre and Tezuka, Masaki",
    title = "{Black Holes and Random Matrices}",
    eprint = "1611.04650",
    archivePrefix = "arXiv",
    primaryClass = "hep-th",
    reportNumber = "SU-ITP-16-19, SU-ITP-16/19, YITP-16-124",
    doi = "10.1007/JHEP05(2017)118",
    journal = "JHEP",
    volume = "05",
    pages = "118",
    year = "2017",
    note = {[Erratum: JHEP 09, 002 (2018)], \href{https://arxiv.org/abs/1611.04650}{\texttt{[1611.04650]}}}
}

@article{Gu_2020,
    author = "Gu, Yingfei and Kitaev, Alexei and Sachdev, Subir and Tarnopolsky, Grigory",
    title = "{Notes on the complex Sachdev-Ye-Kitaev model}",
    eprint = "1910.14099",
    archivePrefix = "arXiv",
    primaryClass = "hep-th",
    doi = "10.1007/JHEP02(2020)157",
    journal = "JHEP",
    volume = "02",
    pages = "157",
    year = "2020",
   note={\href{https://arxiv.org/abs/1910.14099}{\texttt{[1910.14099]}}}
}

@article{Pethybridge_2024,
   title={{Notes on Complex $q=2$ SYK}},
   author={Benjamin James Pethybridge},
   year={2024},
   eprint={2403.04673},
   archivePrefix={arXiv},
   primaryClass={hep-th},
   note={\href{https://arxiv.org/abs/2403.04673}{\texttt{[2403.04673]}}}
}

@article{Zhang_2025,
    author = "Zhang, Ziruo and Peng, Cheng",
    title = "{Gauging the complex SYK model}",
    eprint = "2502.18595",
    archivePrefix = "arXiv",
    primaryClass = "hep-th",
    reportNumber = "USTC-ICTS/PCFT-25-06",
    doi = "10.1007/JHEP08(2025)217",
    journal = "JHEP",
    volume = "08",
    pages = "217",
    year = "2025",
   note={\href{https://arxiv.org/abs/2502.18595}{\texttt{[2502.18595]}}}
}

@article{Fu_2017,
    author = "Fu, Wenbo and Gaiotto, Davide and Maldacena, Juan and Sachdev, Subir",
    title = "{Supersymmetric Sachdev-Ye-Kitaev models}",
    eprint = "1610.08917",
    archivePrefix = "arXiv",
    primaryClass = "hep-th",
    doi = "10.1103/PhysRevD.95.026009",
    journal = "Phys. Rev. D",
    volume = "95",
    number = "2",
    pages = "026009",
    year = "2017",
    note = {[Addendum: Phys.Rev.D 95, 069904 (2017)],
     \href{https://arxiv.org/abs/1610.08917}{\texttt{[1610.08917]}}}
}

@article{Peng_2021,
    author = "Peng, Cheng and Stanojevic, Stefan",
    title = "{Soft modes in $\mathcal{N} = 2$ SYK model}",
    eprint = "2006.13961",
    archivePrefix = "arXiv",
    primaryClass = "hep-th",
    doi = "10.1007/JHEP01(2021)082",
    journal = "JHEP",
    volume = "01",
    pages = "082",
    year = "2021",
   note={\href{https://arxiv.org/abs/2006.13961}{\texttt{[2006.13961]}}}
}

@article{Peng_2017,
    author = "Peng, Cheng and Spradlin, Marcus and Volovich, Anastasia",
    title = "{Correlators in the $\mathcal{N}=2$ Supersymmetric SYK Model}",
    eprint = "1706.06078",
    archivePrefix = "arXiv",
    primaryClass = "hep-th",
    reportNumber = "BROWN-HET-1716",
    doi = "10.1007/JHEP10(2017)202",
    journal = "JHEP",
    volume = "10",
    pages = "202",
    year = "2017",
   note={\href{https://arxiv.org/abs/1706.06078}{\texttt{[1706.06078]}}}
}

@article{Sachdev_2015,
    author = "Sachdev, Subir",
    title = "{Bekenstein-Hawking Entropy and Strange Metals}",
    eprint = "1506.05111",
    archivePrefix = "arXiv",
    primaryClass = "hep-th",
    doi = "10.1103/PhysRevX.5.041025",
    journal = "Phys. Rev. X",
    volume = "5",
    number = "4",
    pages = "041025",
    year = "2015",
   note={\href{https://arxiv.org/abs/1506.05111}{\texttt{[1506.05111]}}}
}

@article{Rosenhaus_2019,
    author = "Rosenhaus, Vladimir",
    title = "{An introduction to the SYK model}",
    eprint = "1807.03334",
    archivePrefix = "arXiv",
    primaryClass = "hep-th",
    doi = "10.1088/1751-8121/ab2ce1",
    journal = "J. Phys. A",
    volume = "52",
    pages = "323001",
    year = "2019",
   note={\href{https://arxiv.org/abs/1807.03334}{\texttt{[1807.03334]}}}
}

@article{Rabinovici_2021,
    author = "Rabinovici, E. and S{\'a}nchez-Garrido, A. and Shir, R. and Sonner, J.",
    title = "{Operator complexity: a journey to the edge of Krylov space}",
    eprint = "2009.01862",
    archivePrefix = "arXiv",
    primaryClass = "hep-th",
    doi = "10.1007/JHEP06(2021)062",
    journal = "JHEP",
    volume = "06",
    pages = "062",
    year = "2021",
   note={\href{https://arxiv.org/abs/2009.01862}{\texttt{[2009.01862]}}}
}

@article{Jian_2021,
    author = "Jian, Shao-Kai and Swingle, Brian and Xian, Zhuo-Yu",
    title = "{Complexity growth of operators in the SYK model and in JT gravity}",
    eprint = "2008.12274",
    archivePrefix = "arXiv",
    primaryClass = "hep-th",
    doi = "10.1007/JHEP03(2021)014",
    journal = "JHEP",
    volume = "03",
    pages = "014",
    year = "2021",
   note={\href{https://arxiv.org/abs/2008.12274}{\texttt{[2008.12274]}}}
}

@article{Heydeman_2024,
    author = "Heydeman, Matthew and Liu, Yifei and Turiaci, Gustavo J.",
    title = "{Supersymmetry breaking in SYK and the black hole spectrum}",
    eprint = "2408.12138",
    archivePrefix = "arXiv",
    primaryClass = "hep-th",
    doi = "10.1007/JHEP06(2025)158",
    journal = "JHEP",
    volume = "06",
    pages = "158",
    year = "2025",
   note={\href{https://arxiv.org/abs/2408.12138}{\texttt{[2408.12138]}}}
}

@book{Viswanath_2013,
  title={{The Recursion Method: Application to Many-Body Dynamics}},
  author={Viswanath, V.S. and M{\"u}ller, G.},
  series={Lecture Notes in Physics Monographs},
  year={1994},
  publisher={Springer Berlin Heidelberg}
}

@article{Roberts_2018,
   author = "Roberts, Daniel A. and Stanford, Douglas and Streicher, Alexandre",
    title = "{Operator growth in the SYK model}",
    eprint = "1802.02633",
    archivePrefix = "arXiv",
    primaryClass = "hep-th",
    doi = "10.1007/JHEP06(2018)122",
    journal = "JHEP",
    volume = "06",
    pages = "122",
    year = "2018",
   note={\href{https://arxiv.org/abs/1802.02633}{\texttt{[1802.02633]}}}
}

@article{Mertens_2023,
    author = "Mertens, Thomas G. and Turiaci, Gustavo J.",
    title = "{Solvable models of quantum black holes: a review on Jackiw{\textendash}Teitelboim gravity}",
    eprint = "2210.10846",
    archivePrefix = "arXiv",
    primaryClass = "hep-th",
    doi = "10.1007/s41114-023-00046-1",
    journal = "Living Rev. Rel.",
    volume = "26",
    number = "1",
    pages = "4",
    year = "2023",
   note={\href{https://arxiv.org/abs/2210.10846}{\texttt{[2210.10846]}}}
}

@article{Saad_2019,
    author = "Saad, Phil and Shenker, Stephen H. and Stanford, Douglas",
    title = "{JT gravity as a matrix integral}",
    eprint = "1903.11115",
    archivePrefix = "arXiv",
    primaryClass = "hep-th",
    month = "3",
    year = "2019",
   note={\href{https://arxiv.org/abs/1903.11115}{\texttt{[1903.11115]}}}
}

@article{Turiaci_2024,
    author = "Turiaci, Gustavo Joaquin",
    title = "{Les Houches lectures on two-dimensional gravity and holography}",
    eprint = "2412.09537",
    archivePrefix = "arXiv",
    primaryClass = "hep-th",
    doi = "10.21468/SciPostPhysLectNotes.113",
    journal = "SciPost Phys. Lect. Notes",
    volume = "113",
    pages = "1",
    year = "2026",
   note={\href{https://arxiv.org/abs/2412.09537}{\texttt{[2412.09537]}}}
}

@article{Caputa_2024,
    author = "Caputa, Pawel and Chen, Bowen and McDonald, Ross W. and Sim{\'o}n, Joan and Strittmatter, Benjamin",
    title = "{Spread complexity rate as proper momentum}",
    eprint = "2410.23334",
    archivePrefix = "arXiv",
    primaryClass = "hep-th",
    reportNumber = "YITP-24-137",
    doi = "10.1103/7zs8-9zpg",
    journal = "Phys. Rev. D",
    volume = "113",
    number = "4",
    pages = "L041901",
    year = "2026",
   note={\href{https://arxiv.org/abs/2410.23334}{\texttt{[2410.23334]}}}
}

@article{Berkooz_2024,
    author = "Berkooz, Micha and Mamroud, Ohad",
    title = "{A cordial introduction to double scaled SYK}",
    eprint = "2407.09396",
    archivePrefix = "arXiv",
    primaryClass = "hep-th",
    doi = "10.1088/1361-6633/ada889",
    journal = "Rept. Prog. Phys.",
    volume = "88",
    number = "3",
    pages = "036001",
    year = "2025",
   note={\href{https://arxiv.org/abs/2407.09396}{\texttt{[2407.09396]}}}
}

@article{Berkooz_2020,
    author = "Berkooz, Micha and Brukner, Nadav and Narovlansky, Vladimir and Raz, Amir",
    title = "{The double scaled limit of Super--Symmetric SYK models}",
    eprint = "2003.04405",
    archivePrefix = "arXiv",
    primaryClass = "hep-th",
    doi = "10.1007/JHEP12(2020)110",
    journal = "JHEP",
    volume = "12",
    pages = "110",
    year = "2020",
   note={\href{https://arxiv.org/abs/2003.04405}{\texttt{[2003.04405]}}}
}

@article{Kanazawa_2017,
    author = "Kanazawa, Takuya and Wettig, Tilo",
    title = "{Complete random matrix classification of SYK models with $\mathcal{N}=0$, $1$ and $2$ supersymmetry}",
    eprint = "1706.03044",
    archivePrefix = "arXiv",
    primaryClass = "hep-th",
    reportNumber = "RIKEN-QHP-311, RIKEN-QHP-311",
    doi = "10.1007/JHEP09(2017)050",
    journal = "JHEP",
    volume = "09",
    pages = "050",
    year = "2017",
   note={\href{https://arxiv.org/abs/1706.03044}{\texttt{[1706.03044]}}}
}

@article{He_2022,
    author = "He, Song and Lau, Pak Hang Chris and Xian, Zhuo-Yu and Zhao, Long",
    title = "{Quantum chaos, scrambling and operator growth in $ T\overline{T} $ deformed SYK models}",
    eprint = "2209.14936",
    archivePrefix = "arXiv",
    primaryClass = "hep-th",
    doi = "10.1007/JHEP12(2022)070",
    journal = "JHEP",
    volume = "12",
    pages = "070",
    year = "2022",
   note={\href{https://arxiv.org/abs/2209.14936}{\texttt{[2209.14936]}}}
}

@article{Kravtsov_2012,
   title={{Random Matrix Theory: Wigner-Dyson Statistics and Beyond}},
   author={V. E. Kravtsov},
   year={2012},
   eprint={0911.0639},
   archivePrefix={arXiv},
   primaryClass={cond-mat.dis-nn},
   note={\href{https://arxiv.org/abs/0911.0639}{\texttt{[0911.0639]}}}
}

@book{Livan_2018,
   title={{Introduction to Random Matrices}},
   journal={Springer Briefs in Mathematical Physics},
   publisher={Springer International Publishing},
   author={Livan, Giacomo and Novaes, Marcel and Vivo, Pierpaolo},
   year={2018}
}

@article{Caputa_2021,
    author = "Caputa, Pawel and Magan, Javier M. and Patramanis, Dimitrios",
    title = "{Geometry of Krylov complexity}",
    eprint = "2109.03824",
    archivePrefix = "arXiv",
    primaryClass = "hep-th",
    doi = "10.1103/PhysRevResearch.4.013041",
    journal = "Phys. Rev. Res.",
    volume = "4",
    number = "1",
    pages = "013041",
    year = "2022",
   note={\href{https://arxiv.org/abs/2109.03824}{\texttt{[2109.03824]}}}
}

@article{Rabinovici_2025,
    author = "Rabinovici, Eliezer and S{\'a}nchez-Garrido, Adri{\'a}n and Shir, Ruth and Sonner, Julian",
    title = "{Krylov Complexity}",
    eprint = "2507.06286",
    archivePrefix = "arXiv",
    primaryClass = "hep-th",
    reportNumber = "CERN-TH-2025-128",
    month = "7",
    year = "2025",
   note={\href{https://arxiv.org/abs/2507.06286}{\texttt{[2507.06286]}}}
}

@article{Tsuji_2018,
    author = "Tsuji, Naoto and Shitara, Tomohiro and Ueda, Masahito",
    title = "{Bound on the exponential growth rate of out-of-time-ordered correlators}",
    eprint = "1706.09160",
    archivePrefix = "arXiv",
    primaryClass = "cond-mat.stat-mech",
    doi = "10.1103/PhysRevE.98.012216",
    journal = "Phys. Rev. E",
    volume = "98",
    pages = "012216",
    year = "2018",
   note={\href{https://arxiv.org/abs/1706.09160}{\texttt{[1706.09160]}}}
}

@book{Cullum_2002,
   author = {Cullum, Jane K. and Willoughby, Ralph A.},
   title = {Lanczos Algorithms for Large Symmetric Eigenvalue Computations},
   publisher = {Society for Industrial and Applied Mathematics},
   year = {2002},
   eprint = {https://epubs.siam.org/doi/pdf/10.1137/1.9780898719192}
}

@article{Iliesiu_2021,
  author = "Iliesiu, Luca V. and Turiaci, Gustavo J.",
    title = "{The statistical mechanics of near-extremal black holes}",
    eprint = "2003.02860",
    archivePrefix = "arXiv",
    primaryClass = "hep-th",
    doi = "10.1007/JHEP05(2021)145",
    journal = "JHEP",
    volume = "05",
    pages = "145",
    year = "2021",
   note={\href{https://arxiv.org/abs/2003.02860}{\texttt{[2003.02860]}}}
}

@article{Heydeman_2021,
    author = "Heydeman, Matthew and Iliesiu, Luca V. and Turiaci, Gustavo J. and Zhao, Wenli",
    title = "{The statistical mechanics of near-BPS black holes}",
    eprint = "2011.01953",
    archivePrefix = "arXiv",
    primaryClass = "hep-th",
    reportNumber = "PUPT-2621",
    doi = "10.1088/1751-8121/ac3be9",
    journal = "J. Phys. A",
    volume = "55",
    number = "1",
    pages = "014004",
    year = "2022",
   note={\href{https://arxiv.org/abs/2011.01953}{\texttt{[2011.01953]}}}
}

@article{Anninos_2023,
    author = "Anninos, Dionysios and Galante, Dami{\'a}n A. and Sheorey, Sameer U.",
    title = "{Renormalisation group flows of deformed SYK models}",
    eprint = "2212.04944",
    archivePrefix = "arXiv",
    primaryClass = "hep-th",
    doi = "10.1007/JHEP11(2023)197",
    journal = "JHEP",
    volume = "11",
    pages = "197",
    year = "2023",
   note={\href{https://arxiv.org/abs/2212.04944}{\texttt{[2212.04944]}}}
}

@article{Chapman_2024,
    author = "Chapman, Shira and Demulder, Saskia and Galante, Dami{\'a}n A. and Sheorey, Sameer U. and Shoval, Osher",
    title = "{Krylov complexity and chaos in deformed Sachdev-Ye-Kitaev models}",
    eprint = "2407.09604",
    archivePrefix = "arXiv",
    primaryClass = "hep-th",
    doi = "10.1103/PhysRevB.111.035141",
    journal = "Phys. Rev. B",
    volume = "111",
    number = "3",
    pages = "035141",
    year = "2025",
   note={\href{https://arxiv.org/abs/2407.09604}{\texttt{[2407.09604]}}}
}

@article{Barbon_2019,
    author = "Barb{\'o}n, J. L. F. and Rabinovici, E. and Shir, R. and Sinha, R.",
    title = "{On The Evolution Of Operator Complexity Beyond Scrambling}",
    eprint = "1907.05393",
    archivePrefix = "arXiv",
    primaryClass = "hep-th",
    reportNumber = "IFT-UAM/CSIC-19-98",
    doi = "10.1007/JHEP10(2019)264",
    journal = "JHEP",
    volume = "10",
    pages = "264",
    year = "2019",
   note={\href{https://arxiv.org/abs/1907.05393}{\texttt{[1907.05393]}}}
}

@article{Guo_2022,
   title={{Operator Growth in SU(2) Yang-Mills Theory}},
   author={Shiyong Guo},
   year={2022},
   eprint={2208.13362},
   archivePrefix={arXiv},
   primaryClass={hep-th},
   note={\href{https://arxiv.org/abs/2208.13362}{\texttt{[2208.13362]}}}
}

@article{Caputa_2025,
    author = "Caputa, Pawel and Di Giulio, Giuseppe and Loc, Tran Quang",
    title = "{Growth of block-diagonal operators and symmetry-resolved Krylov complexity}",
    eprint = "2507.02033",
    archivePrefix = "arXiv",
    primaryClass = "hep-th",
    reportNumber = "YITP-25-101",
    doi = "10.1103/9v9v-54zv",
    journal = "Phys. Rev. Res.",
    volume = "7",
    number = "4",
    pages = "043055",
    year = "2025",
   note={\href{https://arxiv.org/abs/2507.02033}{\texttt{[2507.02033]}}}
}

@article{Caputa_2025_2,
    author = "Caputa, Pawel and Di Giulio, Giuseppe and Loc, Tran Quang",
    title = "{Symmetry-resolved spread complexity}",
    eprint = "2509.12992",
    archivePrefix = "arXiv",
    primaryClass = "hep-th",
    reportNumber = "YITP-25-146",
    doi = "10.1007/JHEP02(2026)189",
    journal = "JHEP",
    volume = "02",
    pages = "189",
    year = "2026",
   note={\href{https://arxiv.org/abs/2509.12992}{\texttt{[2509.12992]}}}
}

@article{Baiguera_2025,
    author = "Baiguera, Stefano and Balasubramanian, Vijay and Caputa, Pawel and Chapman, Shira and Haferkamp, Jonas and Heller, Michal P. and Halpern, Nicole Yunger",
    title = "{Quantum complexity in gravity, quantum field theory, and quantum information science}",
    eprint = "2503.10753",
    archivePrefix = "arXiv",
    primaryClass = "hep-th",
    reportNumber = "YITP-25-39",
    doi = "10.1016/j.physrep.2025.11.001",
    journal = "Phys. Rept.",
    volume = "1159",
    pages = "1--77",
    year = "2026",
   note={\href{https://arxiv.org/abs/2503.10753}{\texttt{[2503.10753]}}}
}

@article{Heydeman_2023,
    author = "Heydeman, Matthew and Turiaci, Gustavo J. and Zhao, Wenli",
    title = "{Phases of $ \mathcal{N} $ = 2 Sachdev-Ye-Kitaev models}",
    eprint = "2206.14900",
    archivePrefix = "arXiv",
    primaryClass = "hep-th",
    doi = "10.1007/JHEP01(2023)098",
    journal = "JHEP",
    volume = "01",
    pages = "098",
    year = "2023",
   note={\href{https://arxiv.org/abs/2206.14900}{\texttt{[2206.14900]}}}
}

@article{Ambrosini_2025_2,
    author = "Ambrosini, Marco and Rabinovici, Eliezer and Sonner, Julian",
    title = "{Holography of K-complexity: switchbacks and shockwaves}",
    eprint = "2510.17975",
    archivePrefix = "arXiv",
    primaryClass = "hep-th",
    reportNumber = "CERN-TH-2025-206",
    doi = "10.1007/JHEP06(2026)060",
    journal = "JHEP",
    volume = "06",
    pages = "060",
    year = "2026",
   note={\href{https://arxiv.org/abs/2510.17975}{\texttt{[2510.17975]}}}
}

@article{Kar_2022,
 author = "Kar, Arjun and Lamprou, Lampros and Rozali, Moshe and Sully, James",
    title = "{Random matrix theory for complexity growth and black hole interiors}",
    eprint = "2106.02046",
    archivePrefix = "arXiv",
    primaryClass = "hep-th",
    doi = "10.1007/JHEP01(2022)016",
    journal = "JHEP",
    volume = "01",
    pages = "016",
    year = "2022",
   note={\href{https://arxiv.org/abs/2106.02046}{\texttt{[2106.02046]}}}
}

@book{Liesen_2012,
  author={Liesen, Jörg and Strakos, Zdenek},
  publisher={Oxford University Press},
  refid={902717176},
  series={Numerical Mathematics and Scientific Computation},
  title={{Krylov Subspace Methods: Principles and Analysis}},
  year=2012
}

@article{Rabinovici_2022,
    author = "Rabinovici, E. and S{\'a}nchez-Garrido, A. and Shir, R. and Sonner, J.",
    title = "{Krylov complexity from integrability to chaos}",
    eprint = "2207.07701",
    archivePrefix = "arXiv",
    primaryClass = "hep-th",
    doi = "10.1007/JHEP07(2022)151",
    journal = "JHEP",
    volume = "07",
    pages = "151",
    year = "2022",
   note={\href{https://arxiv.org/abs/2207.07701}{\texttt{[2207.07701]}}}
}

@article{Rabinovici_2022_2,
    author = "Rabinovici, E. and S{\'a}nchez-Garrido, A. and Shir, R. and Sonner, J.",
    title = "{Krylov localization and suppression of complexity}",
    eprint = "2112.12128",
    archivePrefix = "arXiv",
    primaryClass = "hep-th",
    doi = "10.1007/JHEP03(2022)211",
    journal = "JHEP",
    volume = "03",
    pages = "211",
    year = "2022",
   note={\href{https://arxiv.org/abs/2112.12128}{\texttt{[2112.12128]}}}
}

@article{Garcia_Garcia_2018,
    author = "Garc{\'\i}a-Garc{\'\i}a, Antonio M. and Loureiro, Bruno and Romero-Berm{\'u}dez, Aurelio and Tezuka, Masaki",
    title = "{Chaotic-Integrable Transition in the Sachdev-Ye-Kitaev Model}",
    eprint = "1707.02197",
    archivePrefix = "arXiv",
    primaryClass = "hep-th",
    doi = "10.1103/PhysRevLett.120.241603",
    journal = "Phys. Rev. Lett.",
    volume = "120",
    number = "24",
    pages = "241603",
    year = "2018",
   note={\href{https://arxiv.org/abs/1707.02197}{\texttt{[1707.02197]}}}
}

@article{Garcia_Garcia_2016,
  author = "Garc{\'\i}a-Garc{\'\i}a, Antonio M. and Verbaarschot, Jacobus J. M.",
    title = "{Spectral and thermodynamic properties of the Sachdev-Ye-Kitaev model}",
    eprint = "1610.03816",
    archivePrefix = "arXiv",
    primaryClass = "hep-th",
    doi = "10.1103/PhysRevD.94.126010",
    journal = "Phys. Rev. D",
    volume = "94",
    number = "12",
    pages = "126010",
    year = "2016",
   note={\href{https://arxiv.org/abs/1610.03816}{\texttt{[1610.03816]}}}
}

@article{Li_2017,
    author = "Li, Tianlin and Liu, Junyu and Xin, Yuan and Zhou, Yehao",
    title = "{Supersymmetric SYK model and random matrix theory}",
    eprint = "1702.01738",
    archivePrefix = "arXiv",
    primaryClass = "hep-th",
    doi = "10.1007/JHEP06(2017)111",
    journal = "JHEP",
    volume = "06",
    pages = "111",
    year = "2017",
   note={\href{https://arxiv.org/abs/1702.01738}{\texttt{[1702.01738]}}}
}

@article{Jensen_2016,
    author = "Jensen, Kristan",
    title = "{Chaos in AdS$_2$ Holography}",
    eprint = "1605.06098",
    archivePrefix = "arXiv",
    primaryClass = "hep-th",
    doi = "10.1103/PhysRevLett.117.111601",
    journal = "Phys. Rev. Lett.",
    volume = "117",
    number = "11",
    pages = "111601",
    year = "2016",
   note={\href{https://arxiv.org/abs/1605.06098}{\texttt{[1605.06098]}}}
}

@article{Hamdan2025,
    author = "Abou Hamdan, Weam and Galante, Dami{\'a}n A.",
    title = "{Exploring the infrared landscape of the SYK model}",
    eprint = "2511.14839",
    archivePrefix = "arXiv",
    primaryClass = "hep-th",
    doi = "10.1007/JHEP05(2026)084",
    journal = "JHEP",
    volume = "05",
    pages = "084",
    year = "2026",
   note={\href{https://arxiv.org/abs/2511.14839}{\texttt{[2511.14839]}}}
}

@article{Chang2024,
    author = "Chang, Chi-Ming and Chen, Yiming and Sia, Bik Soon and Yang, Zhenbin",
    title = "{Fortuity in SYK models}",
    eprint = "2412.06902",
    archivePrefix = "arXiv",
    primaryClass = "hep-th",
    doi = "10.1007/JHEP08(2025)003",
    journal = "JHEP",
    volume = "08",
    pages = "003",
    year = "2025",
   note={\href{https://arxiv.org/abs/2412.06902}{\texttt{[2412.06902]}}}
}

@article{Chen2024,
    author = "Chen, Yiming and Lin, Henry W. and Shenker, Stephen H.",
    title = "{BPS chaos}",
    eprint = "2407.19387",
    archivePrefix = "arXiv",
    primaryClass = "hep-th",
    doi = "10.21468/SciPostPhys.18.2.072",
    journal = "SciPost Phys.",
    volume = "18",
    number = "2",
    pages = "072",
    year = "2025",
   note={\href{https://arxiv.org/abs/2407.19387}{\texttt{[2407.19387]}}}
}

@article{Craps2024,
    author = "Craps, Ben and Evnin, Oleg and Pascuzzi, Gabriele",
    title = "{A Relation between Krylov and Nielsen Complexity}",
    eprint = "2311.18401",
    archivePrefix = "arXiv",
    primaryClass = "quant-ph",
    doi = "10.1103/PhysRevLett.132.160402",
    journal = "Phys. Rev. Lett.",
    volume = "132",
    number = "16",
    pages = "160402",
    year = "2024",
   note={\href{https://arxiv.org/abs/2311.18401}{\texttt{[2311.18401]}}}
}

@article{Balasubramanian2021,
    author = "Balasubramanian, Vijay and DeCross, Matthew and Kar, Arjun and Li, Yue (Cathy) and Parrikar, Onkar",
    title = "{Complexity growth in integrable and chaotic models}",
    eprint = "2101.02209",
    archivePrefix = "arXiv",
    primaryClass = "hep-th",
    doi = "10.1007/JHEP07(2021)011",
    journal = "JHEP",
    volume = "07",
    pages = "011",
    year = "2021",
   note={\href{https://arxiv.org/abs/2101.02209}{\texttt{[2101.02209]}}}
}

@article{Das2025,
    author = "Das, Rathindra Nath and Demulder, Saskia and Erdmenger, Johanna and Northe, Christian",
    title = "{Spread complexity for the planar limit of holography}",
    eprint = "2412.09673",
    archivePrefix = "arXiv",
    primaryClass = "hep-th",
    doi = "10.1007/JHEP06(2025)166",
    journal = "JHEP",
    volume = "06",
    pages = "166",
    year = "2025",
   note={\href{https://arxiv.org/abs/2412.09673}{\texttt{[2412.09673]}}}
}

@article{Heller2025a,
    author = "Heller, Michal P. and Papalini, Jacopo and Schuhmann, Tim",
    title = "{Krylov Spread Complexity as Holographic Complexity beyond Jackiw-Teitelboim Gravity}",
    eprint = "2412.17785",
    archivePrefix = "arXiv",
    primaryClass = "hep-th",
    doi = "10.1103/spcr-jgm6",
    journal = "Phys. Rev. Lett.",
    volume = "135",
    number = "15",
    pages = "151602",
    year = "2025",
   note={\href{https://arxiv.org/abs/2412.17785}{\texttt{[2412.17785]}}}
}

@article{Heller2025b,
    author = "Heller, Michal P. and Ori, Fabio and Papalini, Jacopo and Schuhmann, Tim and Wang, Meng-Ting",
    title = "{De Sitter holographic complexity from Krylov complexity in DSSYK}",
    eprint = "2510.13986",
    archivePrefix = "arXiv",
    primaryClass = "hep-th",
    month = "10",
    year = "2025",
   note={\href{https://arxiv.org/abs/2510.13986}{\texttt{[2510.13986]}}}
}

@article{Aguilar-Gutierrez2025,
    author = "Aguilar-Gutierrez, Sergio E.",
    title = "{Evolution with(out) time: relational holography {\&} BPS complexity growth in $ \mathcal{N} $ = 2 double-scaled SYK}",
    eprint = "2510.11777",
    archivePrefix = "arXiv",
    primaryClass = "hep-th",
    doi = "10.1007/JHEP02(2026)229",
    journal = "JHEP",
    volume = "02",
    pages = "229",
    year = "2026",
   note={\href{https://arxiv.org/abs/2510.11777}{\texttt{[2510.11777]}}}
}

@article{Aguilar-Gutierrez2025b,
    author = "Aguilar-Gutierrez, Sergio E.",
    title = "{Building the holographic dictionary of the DSSYK from chords, complexity {\&} wormholes with matter}",
    eprint = "2505.22716",
    archivePrefix = "arXiv",
    primaryClass = "hep-th",
    doi = "10.1007/JHEP10(2025)221",
    journal = "JHEP",
    volume = "10",
    pages = "221",
    year = "2025",
   note={\href{https://arxiv.org/abs/2505.22716}{\texttt{[2505.22716]}}}
}

@article{Aguilar-Gutierrez2025c,
    author = "Aguilar-Gutierrez, Sergio E. and Xu, Jiuci",
    title = "{Geometry of chord intertwiner, multiple shocks and switchback in double-scaled SYK}",
    eprint = "2506.19013",
    archivePrefix = "arXiv",
    primaryClass = "hep-th",
    doi = "10.1007/JHEP02(2026)246",
    journal = "JHEP",
    volume = "02",
    pages = "246",
    year = "2026",
   note={\href{https://arxiv.org/abs/2506.19013}{\texttt{[2506.19013]}}}
}

@article{Aguilar-Gutierrez2025d,
    author = "Aguilar-Gutierrez, Sergio E.",
    title = "{Symmetry sectors in chord space and relational holography in the DSSYK. Lessons from branes, wormholes, and de Sitter space}",
    eprint = "2506.21447",
    archivePrefix = "arXiv",
    primaryClass = "hep-th",
    doi = "10.1007/JHEP10(2025)044",
    journal = "JHEP",
    volume = "10",
    pages = "044",
    year = "2025",
   note={\href{https://arxiv.org/abs/2506.21447}{\texttt{[2506.21447]}}}
}

@article{Aguilar-Gutierrez2024,
    author = "Aguilar-Gutierrez, Sergio E.",
    title = "{Towards complexity in de Sitter space from the doubled-scaled Sachdev-Ye-Kitaev model}",
    eprint = "2403.13186",
    archivePrefix = "arXiv",
    primaryClass = "hep-th",
    doi = "10.1007/JHEP10(2024)107",
    journal = "JHEP",
    volume = "10",
    pages = "107",
    year = "2024",
   note={\href{https://arxiv.org/abs/2403.13186}{\texttt{[2403.13186]}}}
}

@article{Garcia_Garcia_2017,
    author = "Garc{\'\i}a-Garc{\'\i}a, Antonio M. and Verbaarschot, Jacobus J. M.",
    title = "{Analytical Spectral Density of the Sachdev-Ye-Kitaev Model at finite N}",
    eprint = "1701.06593",
    archivePrefix = "arXiv",
    primaryClass = "hep-th",
    doi = "10.1103/PhysRevD.96.066012",
    journal = "Phys. Rev. D",
    volume = "96",
    number = "6",
    pages = "066012",
    year = "2017",
   note={\href{https://arxiv.org/abs/1701.06593}{\texttt{[1701.06593]}}}
}

@article{Guhr_1997,
    author = "Guhr, Thomas and Muller-Groeling, Axel and Weidenmuller, Hans A.",
    title = "{Random matrix theories in quantum physics: Common concepts}",
    eprint = "cond-mat/9707301",
    archivePrefix = "arXiv",
    reportNumber = "H-V27-1997",
    doi = "10.1016/S0370-1573(97)00088-4",
    journal = "Phys. Rept.",
    volume = "299",
    pages = "189--425",
    year = "1998",
   note={\href{https://arxiv.org/abs/cond-mat/9707301}{\texttt{[cond-mat/9707301]}}}
}

@article{Atas_2013,
   title={{Distribution of the Ratio of Consecutive Level Spacings in Random Matrix Ensembles}},
   journal={Physical Review Letters},
   publisher={American Physical Society (APS)},
   author={Atas, Y. Y. and Bogomolny, E. and Giraud, O. and Roux, G.},
         year = 2013,
        month = feb,
       volume = {110},
       number = {8},
          eid = {084101},
        pages = {084101},
          doi = {10.1103/PhysRevLett.110.084101},
   note={\href{https://arxiv.org/abs/1212.5611}{\texttt{[1212.5611]}}}
}

@article{Balasubramanian_2022,
    author = "Balasubramanian, Vijay and Caputa, Pawel and Magan, Javier M. and Wu, Qingyue",
    title = "{Quantum chaos and the complexity of spread of states}",
    eprint = "2202.06957",
    archivePrefix = "arXiv",
    primaryClass = "hep-th",
    doi = "10.1103/PhysRevD.106.046007",
    journal = "Phys. Rev. D",
    volume = "106",
    number = "4",
    pages = "046007",
    year = "2022",
   note={\href{https://arxiv.org/abs/2202.06957}{\texttt{[2202.06957]}}}
}

\end{document}